\documentclass{raa}      

\usepackage{graphicx,times}             
\usepackage{natbib}
\usepackage{amssymb,amsmath}
\bibpunct{(}{)}{;}{a}{}{,}
\usepackage{longtable}
\usepackage{lscape}

\usepackage[pagebackref=true]{hyperref}

\def\hii{H\textsc{ii}}
\def\ks{km s$^{-1}$}

\def\s{$^{\prime\prime}$}

\def\cm3{cm$^{-3}$}

\def\2{$^{12}$CO}
\def\3{$^{13}$CO}
\def\8{C$^{18}$O}

\def\cm2{cm$^{-2}$}

\begin{document}

  \title{Early phases of star formation: testing chemical tools
}

   \volnopage{Vol.0 (20xx) No.0, 000--000}      
   \setcounter{page}{1}          

   \author{N. C. Martinez
      \inst{1}
   \and S. Paron
      \inst{1}
   }

   \institute{CONICET-Universidad de Buenos Aires. Instituto de Astronom\'{\i}a y F\'{\i}sica del Espacio,
            Ciudad Universitaria, (C1428EGA) Ciudad Autónoma de Buenos Aires, Argentina\\
\vs\no
   {\small Received 20xx month day; accepted 20xx month day}}

\abstract{ 
The star forming processes strongly influence the ISM chemistry. 
Nowadays, there are available many high-quality databases at millimeter wavelengths. Using them, it is possible to carry out studies that review and deepen previous results. If these studies involve large samples of sources, it is preferred to use direct tools to study the molecular gas.
With the aim of testing these tools such as the use of the HCN/HNC ratio as a thermometer, and the use of H$^{13}$CO$^{+}$, HC$_{3}$N, N$_{2}$H$^{+}$, and C$_{2}$H as ``chemical clocks'', we present a molecular line study towards 55 sources representing massive young stellar objects (MYSOs) at different evolutive stages: infrared dark clouds (IRDCs), high-mass protostellar objects (HMPOs), hot molecular cores (HMCs) and ultracompact \hii~regions (UC\hii).
We found that the use of HCN/HNC ratio as an universal thermometer in the ISM should be taken with care because the HCN optical depth is a big issue that can affect the method. Hence, this tool should be used only after a careful analysis of the HCN spectrum, checking that no line, neither the main nor the hyperfine ones, present absorption features.
We point out that the analysis of the emission of H$^{13}$CO$^{+}$, HC$_{3}$N, N$_{2}$H$^{+}$, and C$_{2}$H could be useful to trace and distinguish regions among IRDCs, HMPOs and HMCs. 
The molecular line widths of these four species increase from the IRDC to the HMC stage, which can be a consequence of the gas dynamics related to the star-forming processes taking place in the molecular clumps.  
Our results do not only contribute with more statistics regarding to probe such chemical tools, useful to obtain information in large samples of sources, but also complement previous works through the analysis on other types of sources.
}

   \authorrunning{N. C. Martinez \& S. Paron}            
   \titlerunning{Molecules and star formation }  

   \maketitle

%

\section{Introduction} \label{sec:intro}

The sites in which the stars form are characterized by a rich and complex chemistry. The smallest gaseous fragments within a molecular cloud, known as hot molecular cores (HMCs), which are related to the formation of massive stars, are the chemically richest regions in the interstellar medium (ISM) (e.g. \citealt{herbst09,bonfand19}). Molecules and chemistry are ubiquitous along all the stages that a forming star goes through, and moreover, the star-forming processes strongly influence the chemistry of such environments \citep{jorgen20}. For instance, as material collapses and becomes ionized by the young massive stars and shocked by jets and outflows, temperature and density can change drastically, leading to the formation and destruction of molecular species. Thus, observing molecular lines and studying their emission and chemistry is important to shed light on the different stages of star formation and to characterize physical and chemical conditions.

To explore the star-forming processes in different environments, one should work with accurate values of the physical properties, such as the case of the kinetic temperature (T$_{\rm K}$). For example, calculating the excitation temperature (T$_{\rm ex}$) of carbon monoxide $^{12}$C$^{16}$O, the T$_{\rm K}$ of a molecular cloud can be roughly estimated if it is assumed that there is a complete thermalization of the lines (T$_{\rm ex}$ = T$_{\rm K}$). 
Ammonia (NH$_{3}$) has been found in different interstellar environments: from dark quiescent clouds, circumstellar envelopes, and early stages of high and low luminosity star formation to planetary atmosphere and external galaxies
(\citealt{martin79,betz79,ho80,lada08,takano13,boegner22}). Estimating the NH$_{3}$ rotational temperature usually results in a reliable indicator of the kinetic temperature.

Hydrogen cyanide (HNC) and isocyanide (HNC) are two of the most simple molecules in the ISM, first detected almost fifty years ago (\citealt{snyder71,snyder72}). These isomers have a linked chemistry, and differences in the spatial distributions in which they lie within a cloud can reflect the gas chemical conditions and the evolution of the star-forming regions \citep{schilke92}. Recently, \citet{hacar20} proposed the HCN-to-HNC integrated intensity
ratio as a direct and efficient thermometer of the ISM with an optimal working range 15 K $\lesssim$ T$_{\rm K}$ $\leq$ 40 K. The authors performed an analysis of such isomers throughout the 
Integral Shape Filament in Orion deriving an empirical correlation between the HCN/HNC ratio and the kinetic temperature T$_{\rm K}$. Based on the analysis of such correlation towards many dense molecular clumps from the MALT90 survey, they proposed that the HCN/HNC thermometer can be extrapolated for the analysis of the ISM in general, particularly in star-forming sites, aiming to explore it towards different regions and sources along the ISM.

Molecular species that emerge and destroy during the birth of stars can be used to track the star-forming processes within molecular clumps and cores (\citealt{stephens,urquhart}). 
Comparisons between column densities and molecular abundance ratios that can be used to estimate the age and mark the evolutionary stages of star-forming regions are known as ``chemical clocks''.
As \citet{sanhueza12} pointed out, only molecules that show differential abundances with
time can be used to evaluate the evolutionary status of a star-forming region. In general, 
chemical clocks have been studied in depth in low-mass star-forming regions, but it has been less developed in the context of high-mass star-forming regions. As shown by \citet{naiping21}, in the context of an analysis of chemical clocks it is important not only studying abundance ratios of such molecules, but also the integrated line intensities, the line widths, among other parameters. For instance, the molecular line widths ($\Delta$v$^{\rm FWHM}$) are related to the gas kinematics of the molecular clump interior, regarding turbulence, outflows, and shocks among other processes \citep{sanhueza12} which can give information about
the evolutive stage of a MYSO. 

Some interesting molecular species for probing physical and chemical properties of star-forming regions are the diazenylium (N$_{2}$H$^{+}$) and ethynyl radical (C$_{2}$H). Both molecules seem to be good tracers of dense gas in the early stages of the star-forming evolution (\citealt{beuther08,sanhueza12,sanhueza13}),  N$_{2}$H$^{+}$ traces cold gas due to its resistance to depletion at low temperatures (\citealt{Li19}), and the latter, additionally can indicate the presence of a photodissociation region (PDR), where UV photons from young and hot massive stars
irradiate acetylene to produce C$_{2}$H \citep{fuente93,nagy15,garcia17}. HCO$^{+}$ and H$^{13}$CO$^{+}$ (formylium species) are usually employed to investigate infall motions and outflow activity \citep{rawlings04,veena18}, and HC$_{3}$N is helpful to explore gas associated with hot molecular cores \citep{bergin96,taniguchi16,duronea19}. 

The mentioned molecules are among the brightest lines, they were called as molecular fingerprints in a study of a large 
sample of molecular clumps \citep{urquhart}. As the authors pointed out, such molecules are able to trace a large range of physical conditions including cold and dense gas (HNC, H$^{13}$CO$^{+}$,
HCN, HN$^{13}$C, H$^{13}$CN), outflows (HCO$^{+}$), early chemistry (C$_{2}$H),
gas associated with protostars and YSOs (HC$_{3}$N, and cyclic
molecules). Thus, the analysis of such molecules  gives us a significant amount of scope to
search for differences in the chemistry as a function of the evolutionary stage of the star formation taking place within molecular clumps.
For instance, \citet{naiping21} studied 31 extended green objects (EGOs) clumps with data from the MALT90 aiming to better understanding the chemical processes that take place in the evolution of massive star formation. They classified the sources and made a molecular line study over 20 massive young stellar objects (MYSOs) and 11 \hii~regions. Through the comparison of integrated intensities, line widths, and column densities, derived from the emission of N$_{2}$H$^{+}$ and C$_{2}$H with those of the formylium (H$^{13}$CO$^{+}$) and cyanoacetylene (HC$_3$N), they suggested that N$_{2}$H$^{+}$ and C$_{2}$H could act as efficient chemical clocks. They found that the N$_{2}$H$^{+}$ and C$_{2}$H column densities decrease from MYSOs 
to \hii~regions, and the [N$_{2}$H$^{+}$]/[H$^{13}$CO$^{+}$] and [C$_{2}$H]/[H$^{13}$CO$^{+}$] abundance ratios also decrease with the evolutionary stage of the EGO clumps. In addition they found that the velocity widths of N$_{2}$H$^{+}$, C$_{2}$H, H$^{13}$CO$^{+}$, and HC$_{3}$N are comparable to each other in MYSOs, while in \hii~regions the velocity widths 
of N$_{2}$H$^{+}$ and C$_{2}$H tend to be narrower than those of H$^{13}$CO$^{+}$ and HC$_{3}$N.

Nowadays, there are available many high-quality databases generated from observations obtained with the most important (sub)millimeter telescopes such as the IRAM 30\,m Telescope and the Atacama Large Millimeter Array (ALMA), among others.
For instance, using this kind of data, it is possible to carry out new chemical studies that, in turn, review and deepen previous results. If these studies involve large samples of sources, it is preferred to use  direct tools to study and probe the molecular gas as those presented in \citet{hacar20} and \citet{naiping21}.

Infrared dark clouds (IRDCs) are massive, dense, and cold clumps that may harbor budding stars, while high mass protostellar objects (HMPOs) are already protostars accreting material from their surroundings; at this stage, both temperature and density increase, but it is thought that at the beginnings, they are chemically poor from an evolutionary point of view. Hot molecular cores (HMCs) are considered hotter sources where the chemistry is prolific as a consequence of embedded and evolved HMPOs \citep{giann17} until they eventually reach the last stage here considered: UC\hii~regions. The stars responsible for the UC\hii~regions have generally finished their accretion process and have begun to ionize the gas around them. All the processes involved in this evolutive path impact on the molecular gas that eventually can be investigated through the emission of molecular lines.  
It is important to highlight that the mentioned phases in the massive star formation may not have well-defined limits, and sometimes one determined source may have some physical conditions overlapping with those of another kind of source (e.g., \citealt{beuther07,gerner14}). Moreover, different kind of sources may be embedded within the same molecular clump. Given this complex scenario, comparative studies with many sources are needed to analyze the involved physics and chemistry.

With the aim of testing the presented tools concerning the use of the HCN/HNC ratio as a thermometer (following the methodology presented in \citealt{hacar20}), and the analysis of H$^{13}$CO$^{+}$, HC$_3$N, N$_{2}$H$^{+}$, and C$_{2}$H (following the methodology presented in \citealt{naiping21}), we present this study towards 55 sources representing MYSOs at 
different evolutive stages as described above. 
After the presentation of the data and the analyzed sources (Sect.\,\ref{data}), the paper is structured as follows: 
a presentation of the results regarding each tested tool (Sect.\,\ref{results}), their respective discussion (Sect.\,\ref{discus}), and a summary of the main results (Sect.\,\ref{concl}).

\section{Data and analyzed sources}
\label{data}

We made use of the catalog `IRAM 30\,m reduced spectra of 59 sources' 
J/A+A/563/A97 in the ViZieR database\footnote{http://cdsarc.u-strasbg.fr/viz-bin/qcat?J/A+A/563/A97} \citep{gerner14}, from which the molecular data were obtained and the sources were selected to perform our study. 
The particular molecular lines analyzed in our work are presented in Table\,\ref{lines}.
These data, obtained by \citet{gerner14}, with the 30\,m IRAM telescope, are single pointing spectra observed toward each source. The data used here correspond to the 86--94 GHz band, which have an angular and spectral resolution of 29\s~and 0.6 \ks, and typical 1-sigma rms values of about 0.03 K. 

\begin{table}
\centering
\caption{Analyzed molecular lines.}
\label{lines}
\begin{tabular}{lcc}
\hline\hline
Molecule    & Transition &  Rest Frequency          \\
            &            &    (GHz)                 \\
\hline
H$^{13}$CO$^{+}$   & 1--0            & 86.7542               \\             
C$_{2}$H\,{\bf *}           & 1--0 3/2--1/2 F=2--1 & 87.3169           \\
HCN                & 1--0 F=1--1         & 88.6304                    \\
                   & 1--0 F=2--1         & 88.6318            \\
                   & 1--0 F=0--1         & 88.6339                   \\
HNC                & 1--0               & 90.6635             \\
HC$_{3}$N          & 10--9              & 90.9790             \\
N$_{2}$H$^{+}$\,{\bf *}     & 1--0 F$_{1}$=2--1 F=3--2                   & 93.1737                     \\
\hline
\multicolumn{3}{l}{ {\bf *} Species with several observed hyperfine lines. It is only}\\
\multicolumn{3}{l}{included the used line of each molecule in this work. }\\
\end{tabular}
\end{table}

The sources, classified as IRDCs, HMPOs,  HMCs, and UC\hii, are presented in Table\,\ref{sources}, where the coordinates and the distances are included. The classification is based on the guidelines introduced by \citet{beuther07} and \citet{yorke07} according to the physical conditions of the evolutionary sequence in the formation of high-mass stars. IRDCs are sources that consist of cold and dense gas 
and dust that emit mainly at (sub)millimeter wavelengths. According to \citet{gerner14}, the IRDCs of this sample consist of starless
IRDCs as well as IRDCs already starting to harbor point sources at $\mu$m-wavelengths.
HMPOs are sources hosting an actively accreting massive protostar(s), which shows an internal emission source at mid-infrared. HMCs are much warmer than HMPOs and
can be distinguished from a chemical point of view. At this stage, the central source(s) heats the surroundings evaporating molecular-rich ices. Finally, the UV radiation from the embedded 
massive protostar(s) ionizes the surrounding gas and gives rise to an UC\hii~region. Even though there might be overlaps among HMPO, HMC, and even UC\hii~stage,
such classification was followed by \citet{gerner14} based on physical quantities, especially the temperature, which rises from IRDCs to HMPOs to UC\hii~regions, and based on the chemistry that can differentiate HMPOs (chemical poorer sources) from HMCs (molecular richer chemistry sources). The sample includes 19 IRDCs, 20 HMPOs, 7 HMCs, and 9 UC\hii~regions.
In addition to the possible overlaps among the type of sources, it is possible that within the beam of the observational data lies more than one type of source. This is the case of HMC034.26, a region hosting several \hii~regions and a hot core. Thus, it is necessary to be careful with any analysis in relation to the evolutionary stages.

The quality of these IRAM data and the variety of sources in the sample included in such a catalog allow us to perform a new analysis of several molecular species in order to probe the molecular gas conditions and the chemistry related to IRDCs, HMPOs, HMCs, and UC\hii~regions with the aim of probing chemical tools.  

Additionally, we used infrared (IR) data to complement the information obtained from the molecular lines with the aim of probing, in these wavelengths, the activity of the star-forming regions.  Data at the Ks band extracted from the UKIRT Infrared Deep Sky Survey (UKIDSS) \citep{law07}, and IRAC-Spitzer 4.5 $\mu$m data obtained from the GLIMPSE survey \citep{church09} were used.

\begin{table}
\label{sources}
\centering
\caption{Sample of analyzed sources.}
\begin{tabular}{lccccccc}
\hline
Name & R.A.(J2000) & Dec.(J2000) &  Distance & Type  & Ks & 4.5 $\mu$m              &   F$_{\rm peak}$(870 $\mu$m) \\
     &             &             &    (kpc)  &       &              &                &        (Jy beam$^{-1}$)     \\ 
\hline
IRDC011.1	& 18 10 28.4 &	-19 22 34 &	3.60	&	IRDC &  points       & point    & 1.76   \\
IRDC028.1	& 18 42 50.3 &  -04 03 20 & 4.80	&	IRDC &  points       & no       &  -    \\
IRDC028.2	& 18 42 52.1 &	-03 59 54 & 4.80	&	IRDC &  points       &  ext.    &  -    \\
IRDC048.6	& 19 21 44.4 &	 13 49 24 &	2.50	&	IRDC &  points       & no       & -     \\
IRDC079.1	& 20 32 22.0 &	 40 20 10 &	1.00	&	IRDC &  points       & points   &  -     \\	
IRDC079.3	& 20 31 57.7 &	 40 18 26 &	1.00	&	IRDC &  no           & points   & -     \\
IRDC18151	& 18 17 50.3 &	-12 07 54 &	3.00	&	IRDC &  ext.         & no       & -     \\
IRDC18182	& 18 21 15.0 &	-14 33 03 &	3.60    &	IRDC &  points       & points   &  0.99   \\
IRDC18223	& 18 25 08.3 &	-12 45 27 &	3.70	&	IRDC &  points       & no        & 1.58 \\
IRDC18306	& 18 33 32.1 &	-08 32 28 &	3.80	&	IRDC &  points       & no      & 0.68 \\
IRDC18308	& 18 33 34.3 &	-08 38 42 &	4.90	&	IRDC &  points       & no      &  -\\
IRDC18310	& 18 33 39.5 &	-08 21 10 &	5.20	&	IRDC &  points       & no      &  0.92  \\
IRDC18337	& 18 36 18.2 &	-07 41 00 &	4.00	&	IRDC &  points       & no     &  0.66 \\
IRDC18385	& 18 41 17.4 &	-05 09 56 &	3.30	&	IRDC &  points       & no      &  0.53 \\
IRDC18437	& 18 46 21.8 &	-02 12 21 &	7.30	&	IRDC &  points       & points  & 0.68 \\
IRDC184.1	& 18 48 02.1 &	-01 53 56 &	6.40	&	IRDC &  points       & no     &  1.06 \\
IRDC184.3	& 18 47 55.8 &	-01 53 34 &	6.00	&	IRDC &  points       & no      & 2.06 \\
IRDC19175	& 19 19 50.7 &	14 01 23  &	1.10	&	IRDC &  points       & no      &  0.48 \\
IRDC20081	& 20 10 13.0 &	27 28 18  &	0.70	&	IRDC &  no           & x       &  - \\
HMPO18089	& 18 11 51.6 &	-17 31 29 &	3.60	&	HMPO &  ext.         & points+ext. &  9.25     \\
HMPO18102	& 18 13 11.3 &	-18 00 03 &	2.70    &	HMPO &  points       & points      &  3.66        \\
HMPO18151	& 18 17 58.1 &	-12 07 26 &	3.00    &   HMPO &  jets         & points+ext. &  -     \\
HMPO18182	& 18 21 09.2 &	-14 31 50 &	4.50	&   HMPO &  diff.        & points+ext. - EGO  & 5.66      \\
HMPO18247	& 18 27 31.7 &	-11 45 56 &	6.70	&	HMPO &  points       & points+ext.        & 1.87             \\
HMPO18264	& 18 29 14.6 &	-11 50 22 &	3.50    &	HMPO &  ext.         & ext. - EGO         & 7.43  \\
HMPO18310	& 18 33 48.1 &	-08 23 50 &	5.20	&	HMPO &  points       & points             & 2.12 \\
HMPO18488	& 18 51 25.6 &	 00 04 07 &	5.40	&	HMPO &  points       & points             &  2.80 \\
HMPO18517	& 18 54 14.4 &	 04 41 40 &	2.90	&	HMPO &  points       & ext.               &  5.26  \\
HMPO18566	& 18 59 10.1 &	 04 12 14 &	6.70	&	HMPO &  diff.        & points+ext.        &  3.83  \\
HMPO19217	& 19 23 58.8 &	 16 57 44 &	10.50   &	HMPO &  diff.        & points+ext.        &  3.06\\
HMPO19410	& 19 43 11.0 &   23 44 10 &  2.10	&	HMPO &  jets         & ext.               &  5.31  \\
HMPO20126	& 20 14 26.0 &	 41 13 32 &  1.70	&	HMPO &  jets         & x                  &   - \\
HMPO20216	& 20 23 23.8 &	 41 17 40 &  1.70   &	HMPO &  jets         & ext.               &    - \\
HMPO20293	& 20 31 12.9 &	 40 03 20 &  1.30	&   HMPO &  diff.        & ext.               &   -\\
HMPO22134	& 22 15 09.1 &   58 49 09 &  2.60	&	HMPO &  diff.        & ext.               &   -  \\
HMPO23033	& 23 05 25.7 &	 60 08 08 &  3.50	&	HMPO &  x            & x                  &    - \\
HMPO23139	& 23 16 10.5 &	 59 55 28 &  4.80	&	HMPO &  x            & x                  &    - \\
HMPO23151	& 23 17 21.0 &	 59 28 49 &  5.70	&	HMPO &  x            & x                  &    - \\
HMPO23545	& 23 57 06.1 &	 65 24 48 &	 0.80	&	HMPO &  x            & x                  &     -\\
HMC009.62	& 18 06 15.2 & 	-20 31 37 &  5.70	&	HMC &   diff.        & ext.               &    12.48 \\
HMC010.47   & 18 08 38.2 & 	-19 51 50 &	 5.80	&	HMC &   points       & points             &    35.01      \\
HMC029.96	& 18 46 04.0 &	-02 39 21 &  7.40	&	HMC &   points       & ext.                &  12.01   \\
HMC031.41	& 18 47 34.2 &	-01 12 45 &	 7.90   &	HMC &   points       & ext.               &    61.68\\
HMC034.26	& 18 53 18.5 &	 01 14 58 &  4.00   &	HMC &   diff.        & ext.              &     55.65 \\
HMC045.47	& 19 14 25.7 &	 11 09 26 &  6.00   &	HMC &   ext.         & ext. - EGO            &   5.49\\
HMC075.78	& 20 21 44.1 &	 37 26 40 &  4.10   &	HMC &   diff.+ext.   & ext.              &   -\\
UCH005.89	& 18 00 30.4 &	-24 04 00 &  2.50   &	UC\hii & ext.          &  ext.   &    25.73 \\
UCH010.10	& 18 05 13.1 &	-19 50 35 &	 4.40   &	UC\hii & ext.          &  points &      0.92    \\
UCH010.30	& 18 08 55.8 &	-20 05 55 &	 6.00   &	UC\hii & diff.         &  ext. - EGO  & 7.67   \\
UCH012.21	& 18 12 39.7 &	-18 24 20 &  13.50	&	UC\hii & points        &  points      & 11.58  \\
UCH013.87	& 18 14 35.8 &	-16 45 43 &	  4.40  &	UC\hii & diff.+ext.    &  ext.        & 5.52   \\
UCH030.54	& 18 46 59.3 &	-02 07 24 &	  6.10	&	UC\hii & points        &  points+ext. & 2.24    \\
UCH035.20	& 19 01 46.4 &   01 13 25 &	  3.20  &   UC\hii & points+diff.  &  x           & - \\
UCH045.12	& 19 13 27.8 &	 10 53 37 &	  6.90  &	UC\hii & jets          &  ext.           &  7.58 \\
UCH045.45	& 19 14 21.3 &	 11 09 14 &   6.00	&	UC\hii & ext.          &  ext.        &   3.90\\
\hline
\end{tabular}
\end{table}

\section{Results}
\label{results}

\subsection{Infrared and submillimeter continuum emission}

Given that jets and outflows affect the star-forming regions' chemistry, we looked for evidence of such processes by using IR data to have additional information on each source, besides its classification, to interpret and complement the molecular analysis.
Data at the Ks band extracted from the UKIDSS were used to search for signs of jets (e.g. \citealt{paron22}) mainly in HMPOs, and also in HMCs and UC\hii~regions. IRAC-Spitzer 4.5 $\mu$m data, obtained from the GLIMPSE survey, were employed to analyze whether the sources present extended 4.5 $\mu$m, likely indicating outflow activity (e.g. \citealt{davis07}). Additionally, each source was checked if it is cataloged as an ``EGO - likely MYSO outflow candidate'' according to the catalog of \citet{cyga08}.

\begin{figure}
    \centering
    \includegraphics[width=7cm]{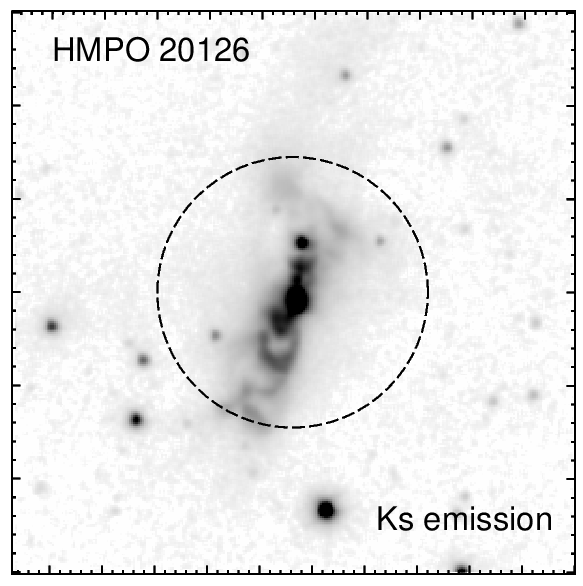}
    \includegraphics[width=7cm]{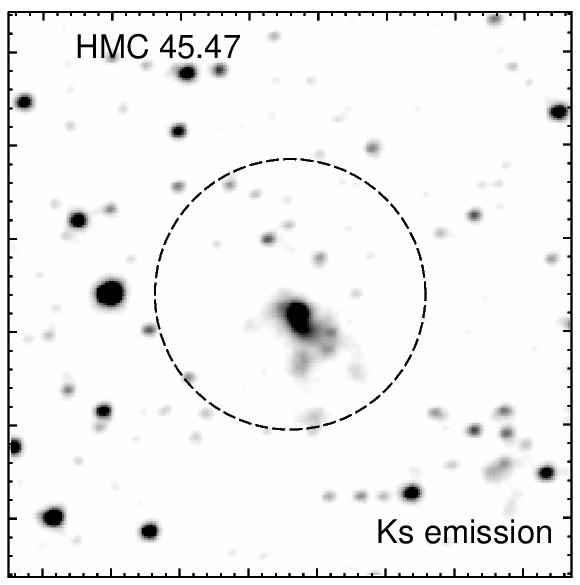}
     \includegraphics[width=7cm]{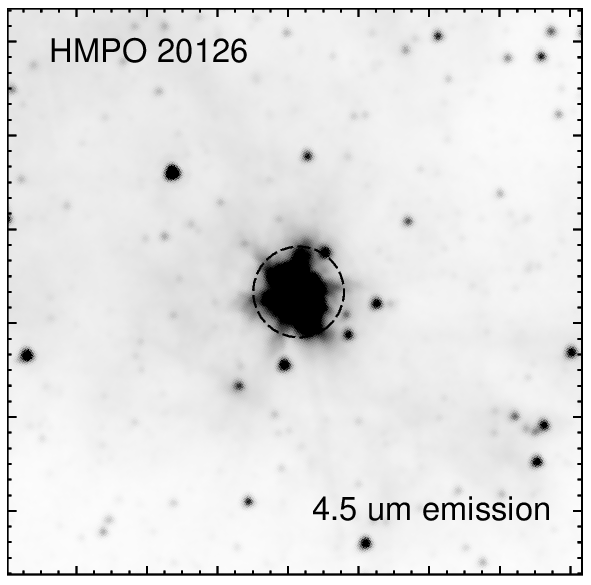} 
     \includegraphics[width=7cm]{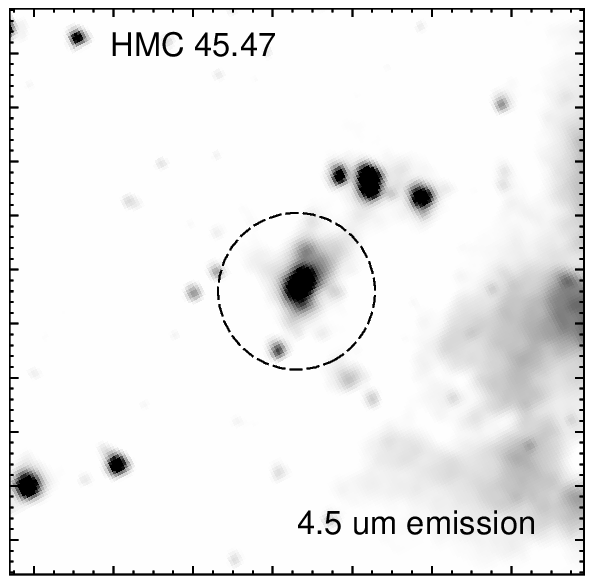}
    \caption{Near-IR Ks emission obtained from the UKIDSS database (top panels) and IRAC-Spitzer 4.5 $\mu$m emission (bottom panels) towards HMPO\,20126 and HMC\,45.47, respectively. In all cases, the dashed circle represents the position and the beam size (29\s) of the IRAM observations.}
    \label{irfig}
\end{figure}

In Table\,\ref{sources} (Cols.\,6 and 7), we indicate the kind of emission at the Ks and 4.5 $\mu$m bands observed within the beam size of the IRAM data centered at the position of each source. If one or several point sources appear within the IRAM beam, it is indicated as `points'; extended emission, suggesting the presence of jets and outflows, is indicated as `ext.', and if this extended emission present a jet-like morphology, it is stated as `jets'. Diffuse emission without any clear morphology is indicated as `diff.'. While `no' means that there is no emission, an `x' indicates there is no data at the corresponding wavelength. In the case of the 4.5 $\mu$m emission, if the source is cataloged as an ``EGO - likely MYSO outflow candidate'', this is indicated with `EGO' in Col.\,7. As an example, Fig.\,\ref{irfig} displays two sources in the Ks band: on the left HMC\,45.47 shows a jet-like feature and some extended emission at the 4.5 $\mu$m band, on the right HMC\,45.47 appears with extended emission at both bands, and it is cataloged as an EGO.

Additionally, in Table\,\ref{sources} (Col.\,8), we include the peak flux of the submillimeter emission obtained from the ATLASGAL compact source catalog \citep{contreras13,urqu14} for sources that lie within the IRAM beam centered at the coordinates indicated in Cols.\,2 and 3.

\subsection{HCN/HNC ratio and T$_{\rm K}$}

Following the same procedure as done by \citet{hacar20} when they test the HCN/HNC kinetic temperature in a large sample of dense molecular clumps 
extracted from the MALT90 survey (see their Sect.\,4.4), we derived the T$_{\rm K}$ for each source of the sample presented here. The correlation to derive T$_{\rm K}$ proposed by the authors is as follows: if the integrated intensity ratio of the isomers HCN and HNC (hereafter I(HCN)/I(HNC)) is $\leq$\, 4, T$_{\rm K}$ is estimated as T$_{\rm K}$ $= 10\times$I(HCN)/I(HNC); and if I(HCN)/I(HNC) $>$ 4, it is used T$_{\rm K} = 3\times$(I(HCN)/I(HNC) $- 4) + 40$. As done by \citet{hacar20}, we integrated the HCN (J=1--0), including all hyperfine components and HNC (J=1--0) over a given velocity interval (in Appendix.\,\ref{appx} we include some spectra indicating the integrated area in both isomers). The obtained integrated emission values, and the respective ratios, are presented in Table\,\ref{resultsI} (Cols.\,2, 3, and 4, respectively), and the obtained kinetic temperatures following the mentioned relations are presented in Table\,\ref{tkresults}. Typical errors in I(HCN) and I(HNC) are about 0.2 and 0.1 K \ks, respectively, which yields errors between 1 and 5\%, and in some cases at most 10\%, in the T$_{\rm K}$. From now on, the obtained errors in the measured and derived parameters appear as bars in the figures, and for a better display, we do not include them in the tables.  

In the further analysis and comparisons, following \citet{hacar20}, we do not consider values of kinetic temperatures obtained from the HCN/HNC ratio (hereafter T$_{\rm K}$(HCN/HNC)) lower than 15 K. These values are due to I(HCN)/I(HNC) ratios close to the unity, which according to the authors, the uncertainties in the method grow up for such cases, likely due to the combination of excitation and opacity effects. In Sect.\,\ref{low15}, based on the inspection of the HCN and HNC spectra of sources that T$_{\rm K}$(HCN/HNC)$<$15 K, we discuss this issue.

In order to compare the T$_{\rm K}$(HCN/HNC), we sought the dust temperature of each source.  Dust temperature ($\rm T_{dust}$) values were obtained from the maps\footnote{http://www.astro.cardiff.ac.uk/research/ViaLactea/} generated by the point process mapping (PPMAP) algorithm \citep{marsh15} done to the Hi-GAL maps in the wavelength range 70–500 $\mu$m \citep{marsh}. $\rm T_{dust}$ values are included in Table\,\ref{tkresults}, and they represent average values on the dust temperature map over the IRAM beam size centered at the source position. Additionally, we also compared with the ammonia kinetic temperature ($\rm T_{K}$(NH$_{3}$)) derived by \citet{urqu11} in several sources of the sample analyzed here that are contained in the Red MSX Source survey (\citealt{hoare05,mottram06,urqu08}). 

\begin{figure*}
    \centering
    \includegraphics[width=7cm]{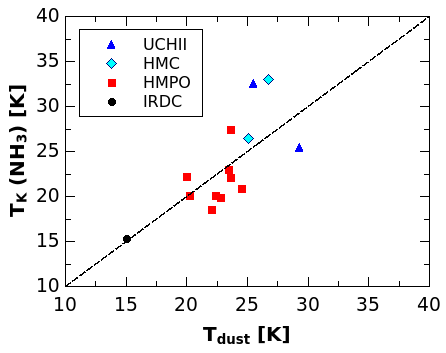}
    \includegraphics[width=7cm]{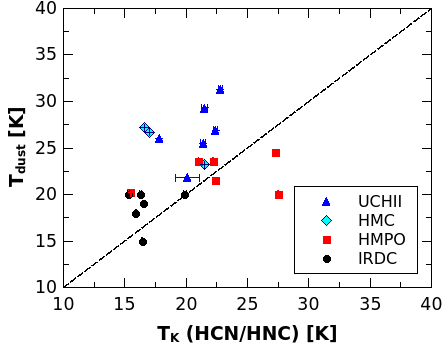}
    \includegraphics[width=7cm]{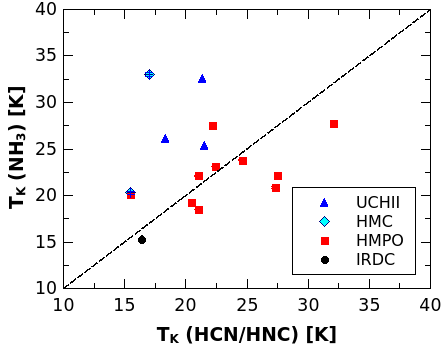}
    \caption{Up-left: Kinetic temperature of ammonia derived from transitions (J,K) = (2,2)-(1,1) versus dust temperature.
    Up-right: Dust temperature obtained from PPMAP maps produced with the Hi-GAL maps \citep{marsh} versus kinetic temperature derived from the
    HCN-HNC integrated intensity ratio. Bottom: Kinetic temperature obtained from the ammonia versus kinetic temperature derived from the HCN-HNC integrated intensity ratio. The dashed black line, in all cases, indicates unity. Error bars in the $\rm T_{K}$(HCN/HNC) are displayed, and most of them are represented by the symbol size.}
    \label{compara}
\end{figure*}

Figure\,\ref{compara} exhibits the comparisons between $\rm T_{K}$(NH$_{3}$) and $\rm T_{dust}$ (upper panel), $\rm T_{dust}$ and  $\rm T_{K}$(HCN/HNC) (middle panel), and $\rm T_{K}$(NH$_{3}$) and $\rm T_{K}$(HCN/HNC) (bottom panel). Table\,\ref{deltaT} presents the average temperature values with their
errors. As mentioned above, sources with values T$_{\rm K}$(HCN/HNC)$<$15 K, were not considered neither in the figures nor in the calculated average. 

\begin{table}
\caption{\label{resultsI}Integrated line intensities (units in K \ks) and the ratio between HCN and HNC.}
\centering
\tiny
\begin{tabular}{lccccccc}
\hline
Source &  I(HCN) & I(HNC) & I(HCN/HNC) &  I(${\rm N_{2}H^{+}}$)$^{a}$ & I(${\rm HC_{3}N}$) &  I(${\rm H^{13}CO^{+}})$ & I(${\rm C_{2}H}$)$^{b}$ \\
\hline
IRDC011.1&6.83 &7.27 &0.94 &8.74 &2.29 &2.01 &1.52 \\
IRDC028.1&16.70 &12.50 &1.34 &10.50 &3.49 &1.83 &2.02 \\ 
IRDC028.2&8.71 &8.34 &1.04 &11.20 &7.62 &2.46 &3.14  \\
IRDC048.6&3.36 &2.80 &1.20 &2.84 &0.22 &0.69 &0.25 \\
IRDC079.1&7.96 &4.86 &1.64 &6.11 &0.69 &2.01 &1.55 \\
IRDC079.3&6.40 &5.50 &1.16 &5.59 &0.97 &2.08 &1.94 \\
IRDC18151&49.10 &19.50 &2.52 &13.90 &2.42 &2.63 &5.12 \\ 
IRDC18182&6.03 &3.93 &1.54 &4.63 &0.74 &0.79 &1.32 \\
IRDC18223&18.90 &11.90 &1.59 &13.20 &2.04 &2.60 &2.11 \\ 
IRDC18306&1.28 &2.45 &0.52 &2.57 &0.40 &0.58 &0.66  \\
IRDC18308&14.30 &8.68 &1.65 &6.51 &1.25 &0.85 &1.34 \\
IRDC18310&2.59 &5.15 &0.50 &9.38 &1.76 &1.02 &0.73 \\
IRDC18337&1.08 &1.80 &0.60 &5.94 &- &0.95 &0.89 \\
IRDC18385&6.51 &3.27 &1.99 &4.79 &1.08 &0.82 &1.03 \\
IRDC18437&6.93 &4.24 &1.63 &-&0.34 &0.40 &1.07 \\
IRDC184.1&0.60 &1.29 &0.47 &4.78 &0.45 &0.60 &1.11 \\
IRDC184.3&11.50 &12.80 &0.90 &- &1.86 &- &2.51  \\
IRDC19175&2.46 &2.54 &0.97 &1.76 &0.18 &0.47 &0.52 \\
IRDC20081&6.62 &2.96 &2.24 &2.11 &-&1.02 &1.20 \\
HMPO18089&10.40 &8.57 &1.21 &17.90 &9.75 &4.80 &5.23 \\ 
HMPO18102&22.60 &19.90 &1.14 &19.70 &5.71 &2.97 &2.87 \\
HMPO18151&49.10 &20.00 &2.46 &11.40 &4.43 &3.50 &5.18 \\
HMPO18182&34.00 &15.20 &2.24 &14.90 &5.55 &3.78 &4.28 \\
HMPO18247&10.60 &8.64 &1.23 &8.29 &1.13 &1.24 &1.83 \\
HMPO18264&74.10 &26.90 &2.76 &25.50 &6.55 &5.08 &7.14 \\
HMPO18310&8.60 &8.52 &1.01 &11.20 &2.41 &2.20 &3.19 \\
HMPO18488&7.31 &10.20 &0.72 &13.20 &3.88 &2.59 &3.93 \\
HMPO18517&48.20 &20.50 &2.35 &14.00 &5.03 &4.85 &6.66 \\
HMPO18566&11.20 &11.60 &0.97 &12.50 &5.72 &3.02 &2.95 \\
HMPO19217&18.30 &9.98 &1.83 &11.60 &3.22 &3.20 &3.10 \\
HMPO19410&38.70 &18.40 &2.10 &22.50 &5.70 &3.17 &6.11 \\
HMPO20126&55.70 &24.90 &2.24 &15.90 &5.68 &3.65 &6.63 \\
HMPO20216&15.90 &7.74 &2.05 &3.34 &0.98 &1.31 &2.56 \\
HMPO20293&29.70 &19.20 &1.55 &21.30 &4.56 &2.55 &5.22 \\
HMPO22134&13.30 &4.87 &2.73 &2.51 &1.82 &1.09 &3.07 \\
HMPO23033&35.70 &17.00 &2.10 &12.80 &3.99 &3.26 &5.97 \\
HMPO23139&31.10 &9.69 &3.21 &6.91 &2.72 &1.49 &4.13 \\
HMPO23151&14.50 &3.38 &4.29 &-&0.35 &0.43 &2.36 \\
HMPO23545&6.79 &2.14 &3.18 &0.39 &0.14 &0.74 &1.28 \\
HMC009.62&45.10 &38.70 &1.17 &19.70 &15.30 &8.91 &11.90 \\ 
HMC010.47&110.30 &51.20 &2.15 &-&24.50 &9.69 &12.60 \\
HMC029.96&49.20 &28.90 &1.70 &10.50 &11.20 &6.10 &6.97  \\
HMC031.41&19.70 &12.70 &1.55 &-&15.40 &2.72 &6.14 \\
HMC034.26&24.70 &42.70 &0.58 &- &27.20 &14.50 &9.90 \\ 
HMC045.47&23.20 &20.90 &1.11 &22.50 &6.72 &5.75 &5.06 \\
HMC075.78&33.90 &20.40 &1.66 &11.00 &6.78 &4.51 &6.36 \\
UCH005.89&130.50 &91.70 &1.42 &35.00 &57.60 &15.40 &21.8 \\
UCH010.10&4.48 &2.23 &2.01 &-&-&-&- \\
UCH010.30&95.10 &53.30 &1.78 &18.50 &19.00 &6.49 &15.10 \\ 
UCH012.21&13.60 &24.10 &0.56 &-&10.50 &4.30 &7.42 \\
UCH013.87&31.30 &14.00 &2.24 &7.67 &4.35 &3.59 &6.90 \\
UCH030.54&21.20 &9.92 &2.14 &2.67 &1.03 &0.93 &2.61 \\
UCH035.20&26.40 &14.40 &1.83 &7.27 &3.13 &2.46 &3.90 \\
UCH045.12&43.40 &19.00 &2.28 &2.97 &2.34 &2.99 &5.42 \\
UCH045.45&22.10 &10.30 &2.15 &4.06 &1.64 &1.28 &2.96 \\
\hline
\multicolumn{8}{l}{$^{a}$ Integrated line intensity of the ${\rm N_{2}H^{+}}$ main component (1–0 F$_{1}$=2–1 F=3–2).}\\
\multicolumn{8}{l}{$^{b}$ Integrated line intensity of the ${\rm C_{2}H}$ main component (1–0 3/2-1/2 F=2–1).}\\
\end{tabular}
\end{table}

\begin{table}
\caption{\label{resultsD}FWHM $\Delta$v from Gaussian fittings (units in \ks).}
\tiny
\centering
\begin{tabular}{lcccc}
\hline
Source &  $\Delta$v(${\rm N_{2}H^{+}}$)$^{a}$ & $\Delta$v(${\rm HC_{3}N}$) &  $\Delta$v(${\rm H^{13}CO^{+}})$ & $\Delta$v(${\rm C_{2}H}$)$^{b}$ \\
\hline
IRDC011.1&3.03 &2.10 &2.11 &2.11 \\
IRDC028.1&3.39 &2.87 &2.69 &2.57 \\
IRDC028.2&2.75 &3.62 &3.22 &3.53 \\
IRDC048.6&2.27 &1.31 &1.52 &1.11 \\
IRDC079.1&2.30 &1.01 &1.68 &1.79 \\
IRDC079.3&2.54 &1.20 &1.68 &1.72 \\
IRDC18151&3.08 &3.28 &2.53 &3.85 \\
IRDC18182&2.76 &2.17 &1.61 &2.45 \\
IRDC18223&3.41 &2.86 &2.78 &2.87 \\
IRDC18306&2.84 &2.43 &1.74 &2.03 \\
IRDC18308&3.82 &2.68 &2.17 &3.00 \\
IRDC18310&3.46 &2.97 &2.11 &2.61 \\
IRDC18337&3.10 &- &1.84 &2.85 \\
IRDC18385&2.67 &3.17 &1.30 &1.96 \\
IRDC18437&- &1.74 &1.52 &1.88  \\
IRDC184.1&2.88 &2.32 &2.53 &2.90 \\
IRDC184.3&- &2.52 &- &4.60 \\
IRDC19175&2.82 &- &1.95 &1.72 \\
IRDC20081&1.91 &-&1.16 &1.56 \\
HMPO18089&3.39 &3.60 &3.54 &3.47 \\ 
HMPO18102&5.00 &3.67 &3.65 &3.49 \\
HMPO18151&2.50 &1.90 &2.16 &2.26 \\
HMPO18182&3.55 &3.05 &3.02 &3.29 \\
HMPO18247&3.15 &2.43 &2.52 &2.66 \\
HMPO18264&3.31 &3.14 &2.53 &3.01 \\
HMPO18310&3.14 &1.88 &2.33 &2.23 \\
HMPO18488&3.48 &3.20 &3.15 &4.61 \\
HMPO18517&3.16 &2.86 &2.52 &2.97 \\
HMPO18566&3.94 &3.62 &3.73 &4.20 \\
HMPO19217&4.41 &4.25 &3.99 &3.75 \\
HMPO19410&2.50 &1.85 &1.80 &1.99 \\
HMPO20126&2.54 &2.41 &2.38 &2.34 \\
HMPO20216&2.20 &1.33 &1.37 &1.82 \\
HMPO20293&2.58 &1.88 &1.94 &2.03 \\
HMPO22134&2.17 &1.25 &1.69 &1.84 \\
HMPO23033&3.00 &2.35 &2.52 &2.58 \\
HMPO23139&2.83 &2.50 &2.48 &2.92 \\
HMPO23151&-&1.81 &2.24 &2.42      \\
HMPO23545&2.05 &1.34 &2.34 &1.80 \\
HMC009.62&4.41 &4.96 &4.02 &4.14 \\
HMC010.47&- &7.48 &7.06 &6.95      \\
HMC029.96&3.09 &3.52 &2.72 &3.27 \\
HMC031.41&- &5.99 &4.90 &7.03       \\
HMC034.26&- &6.33 &4.71 &3.46   \\
HMC045.47&4.41 &3.93 &4.31 &4.33 \\
HMC075.78&3.90 &3.00 &3.54 &4.36 \\
UCH005.89&3.50 &3.94 &3.64 &3.33 \\
UCH010.10&-&-&-&-                  \\
UCH010.30&4.58 &4.54 &4.35 &6.26  \\
UCH012.21&- &7.87 &5.94 &6.14 \\
UCH013.87&3.14 &2.46 &2.53 &2.94 \\ 
UCH030.54&3.59 &2.42 &2.55 &3.18 \\
UCH035.20&3.29 &3.23 &3.32 &2.73 \\
UCH045.12&4.05 &3.98 &3.39 &4.00 \\
UCH045.45&3.20 &3.92 &3.41 &4.70 \\
\hline
\multicolumn{5}{l}{$^{a}$ $\Delta$v of the ${\rm N_{2}H^{+}}$ main component (1–0 F$_{1}$=2–1 F=3–2).}\\
\multicolumn{5}{l}{$^{b}$ $\Delta$v of the ${\rm C_{2}H}$ main component (1–0 3/2-1/2 F=2–1).}\\
\end{tabular}
\end{table}

\begin{table}
\centering
\caption{\label{tkresults}Kinetic temperatures obtained from the different methods (units in K).}
\begin{tabular}{lccc|cccc}
\hline
Source  & T$_{\rm K}$(HCN/HNC) &  T$_{\rm dust}$ & T$_{\rm K }$(NH$_{3}$) & Source  & T$_{\rm K}$(HCN/HNC) &  T$_{\rm dust}$ & T$_{\rm K }$(NH$_{3}$) \\
\hline
IRDC011.1	&	9.4	&	18.0	&	-	&	HMPO18566	&	9.7	&	22.8	&	19.9	\\
IRDC028.1	&	13.4	&	17.0	&	-	&	HMPO19217	&	18.3	&	23.6	&	27.5	\\
IRDC028.2	&	10.4	&	17.0	&	-	&	HMPO19410	&	21.0	&	23.6	&	22.1	\\
IRDC048.6	&	12.0	&	19.0	&	-	&	HMPO20126	&	22.4	&	-	&	23.1	\\
IRDC079.1	&	16.4	&	15.0	&	15.3	&	HMPO20216	&	20.5	&	-	&	19.2	\\
IRDC079.3	&	11.6	&	14.0	&	-	&	HMPO20293	&	15.5	&	20.2	&	20.1	\\
IRDC18151	&	25.2	&	-	&	-	&	HMPO22134	&	27.3	&	24.5	&	20.9	\\
IRDC18182	&	15.3	&	20.0	&	-	&	HMPO23033	&	21.0	&	-	&	18.5	\\
IRDC18223	&	15.9	&	18.0	&	-	&	HMPO23139	&	32.1	&	-	&	27.7	\\
IRDC18306	&	5.2	&	19.0	&	-	&	HMPO23151	&	40.9	&	-	&	-	\\
IRDC18308	&	16.5	&	19.0	&	-	&	HMPO23545	&	31.8	&	-	&	-	\\
IRDC18310	&	5.0	&	20.0	&	-	&		&		&		&		\\
IRDC18337	&	6.0	&	20.0	&	-	&	HMC009.62	&	11.7	&	25.1	&	-	\\
IRDC18385	&	19.9	&	20.0	&	-	&	HMC010.47	&	21.5	&	23.2	&	-	\\
IRDC18437	&	16.3	&	20.0	&	-	&	HMC029.96	&	17	&	26.7	&	33.0	\\
IRDC184.1	&	4.7	&	20.0	&	-	&	HMC031.41	&	15.5	&	-	&	20.3	\\
IRDC184.3	&	9.0	&	19.0	&	-	&	HMC034.26	&	5.8	&	28.4	&	-	\\
IRDC19175	&	9.7	&	20.0	&	-	&	HMC045.47	&	11.1	&	25.1	&	26.4	\\
IRDC20081	&	22.4	&	-	&	-	&	HMC075.78	&	16.6	&	27.2	&	-	\\
	&		&		&		&		&		&		&		\\
HMPO18089	&	12.1	&	23.4	&	23.0	&	UCH005.89	&	14.2	&	26.8	&	-	\\
HMPO18102	&	11.4	&	17.8	&	-	&	UCH010.10	&	20.1	&	21.8	&	-	\\
HMPO18151	&	24.6	&	-	&	23.8	&	UCH010.30	&	17.8	&	26	&	-	\\
HMPO18182	&	22.4	&	21.5	&	-	&	UCH012.21	&	5.6	&	23.1	&	-	\\
HMPO18247	&	12.3	&	22.4	&	20.1	&	UCH013.87	&	22.4	&	26.9	&	-	\\
HMPO18264	&	27.5	&	20	&	22.2	&	UCH030.54	&	21.4	&	25.5	&	32.6	\\
HMPO18310	&	10.1	&	20	&	-	&	UCH035.20	&	18.3	&	-	&	26.1	\\
HMPO18488	&	7.2	&	22	&	18.6	&	UCH045.12	&	22.8	&	31.3	&	-	\\
HMPO18517	&	23.5	&	-	&	-	&	UCH045.45	&	21.5	&	29.3	&	25.4	\\
\hline
\end{tabular}
\end{table}

\begin{table}
\centering
\caption{Calculated average temperatures (units in K).}
\label{deltaT}
\begin{tabular}{lccc}
\hline\hline
Source	&	T${_{\rm dust}}$	&	T$_{\rm K}$(NH$_{3}$) & $\rm T_{K}$(HCN/HNC)		\\
\hline
IRDC	& 18.5$\pm$0.4	&	15.3$^{*}$    &	18.5$\pm$1.2		\\
HMPO	& 21.8$\pm$0.6	&	22.2$\pm$0.8  &	24.9$\pm$1.7		\\
HMC	    & 26.0$\pm$0.6	&	26.6$\pm$3.7  &	16.3$\pm$1.6		\\
UC\hii	& 26.3$\pm$1.1	&	28.0$\pm$2.3  &	19.8$\pm$1.0		\\

\hline
\multicolumn{4}{l}{$^{*}$Value obtained from a single source.}
\end{tabular}
\end{table}

\begin{table}
\centering
\caption{Average values of the molecular line widths (units in \ks).}
\label{deltaprom}
\begin{tabular}{lcccc}

\hline\hline
Source & $\Delta$v(${\rm N_{2}H^{+}}$) & $\Delta$v(${\rm HC_{3}N}$) &  $\Delta$v(${\rm H^{13}CO^{+}}$) & $\Delta$v(${\rm C_{2}H}$) \\
\hline
IRDC	&	2.88$\pm$0.11	&	2.39$\pm$0.19	&	2.00$\pm$0.13	&	2.47$\pm$0.20 	\\
HMPO	&	3.09$\pm$0.17	&	2.51$\pm$0.19	&	2.59$\pm$0.16	&	2.78$\pm$0.18 	\\
HMC	    &	3.95$\pm$0.31	&	5.02$\pm$0.62	&	4.46$\pm$0.51	&	4.79$\pm$0.58	\\
UC\hii	&	3.62$\pm$0.20	&	4.04$\pm$0.60	&	3.64$\pm$0.38	&	4.19$\pm$0.49	\\
\hline
\end{tabular}
\end{table}

\subsection{Molecules as ``chemical clocks''?}

Following \citet{naiping21}, we performed a similar analysis using the HNC, C$_{2}$H, HC$_{3}$N, H$^{13}$CO$^{+}$ and N$_{2}$H$^{+}$ emissions measured toward the sources presented in Table\,\ref{sources}. This is: 1) we compared the molecular emissions with the flux at the sub-millimeter continuum, 2) we analyzed the line widths of the C$_{2}$H, HC$_{3}$N, H$^{13}$CO$^{+}$ and N$_{2}$H$^{+}$, and 3) we analyzed the column densities (extracted from \citealt{gerner14}) of C$_{2}$H and N$_{2}$H$^{+}$ by comparing with that of the H$^{13}$CO$^{+}$.

\begin{figure*}[h!]
    \centering
    \includegraphics[width=7cm]{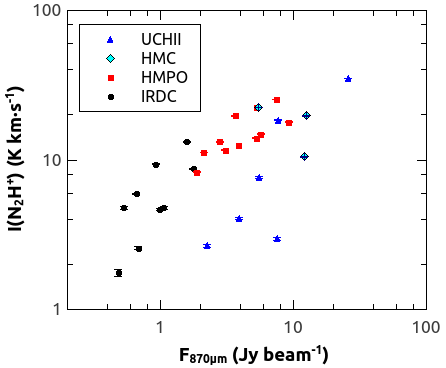}
    \includegraphics[width=7cm]{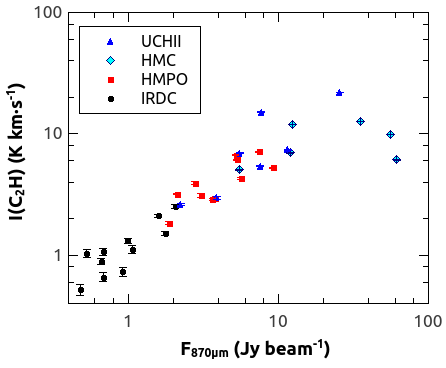}
    \includegraphics[width=7cm]{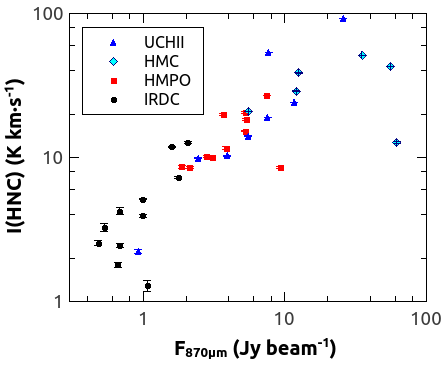}
    \includegraphics[width=7cm]{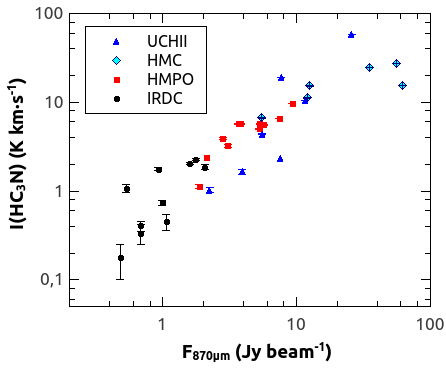}
    \caption{Plots of the integrated line emission versus the 870 $\mu$m peak flux. Top panels display N$_{2}$H$^{+}$ (main component, left) and C$_{2}$H (main component, right). Bottom panels display: HNC (left) and HC$_{3}$N (right). Error bars in the y-axis represent the formal error in the integration of the line emissions.}
    \label{F870}
\end{figure*}

Figure\,\ref{F870} displays the integrated intensities of HNC, C$_{2}$H, HC$_{3}$N and N$_{2}$H$^{+}$ (presented in Table\,\ref{resultsI}) versus the 870 $\mu$m peak flux obtained from the ATLASGAL compact source catalog (presented in Col.\,8 in Table\,\ref{sources}). 

\begin{figure*}[h!]
    \centering
    \includegraphics[width=7cm]{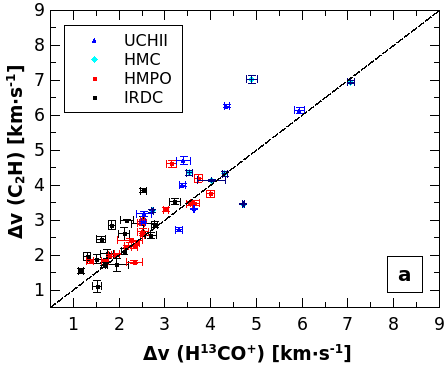}
    \includegraphics[width=7cm]{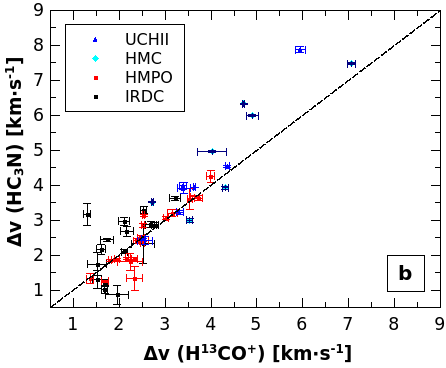}
    \includegraphics[width=7cm]{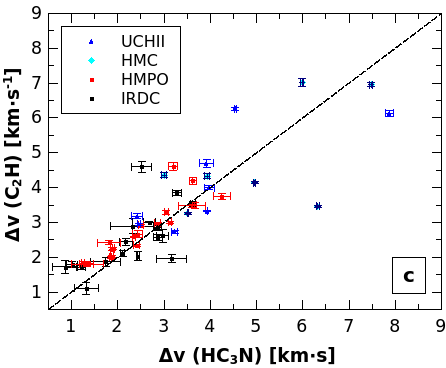}
    \includegraphics[width=7cm]{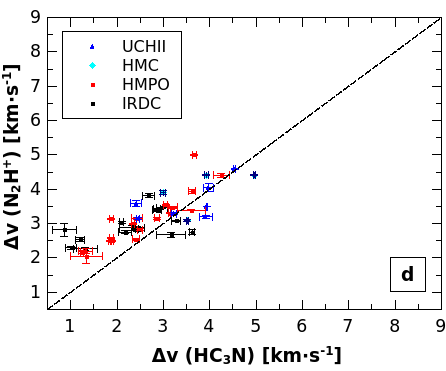}
    \includegraphics[width=7cm]{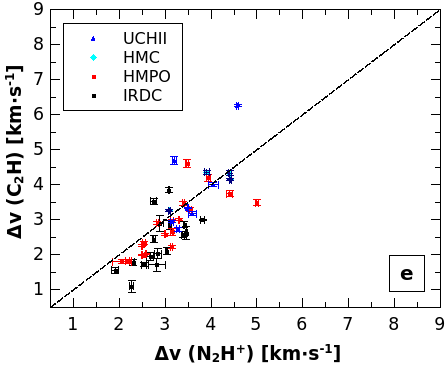}
    \includegraphics[width=7cm]{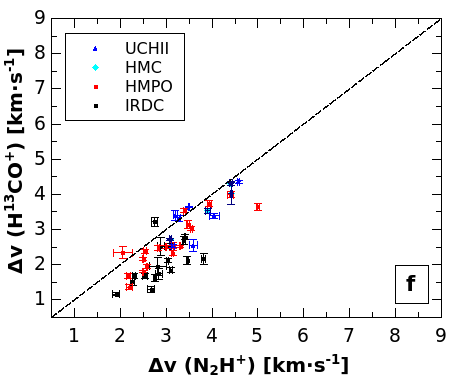}
    \caption{Plots of the FWHM line widths ($\Delta$v) obtained from Gaussian fittings to the emission lines of C$_{2}$H, H$^{13}$CO$^{+}$, HC$_{3}$N, and N$_{2}$H$^{+}$. The dashed black line indicates unity. Error bars represent the formal errors of the Gaussian fittings.}
    \label{deltav}
\end{figure*}

We fitted the line emission of C$_{2}$H, N$_{2}$H$^{+}$, H$^{13}$CO$^{+}$, and HC$_{3}$N with Gaussian functions to obtain the FWHM line widths ($\Delta$v) in each source (in Appendix\,\ref{appx} we include some spectra showing the Gaussian fittings). In the case of the C$_{2}$H and N$_{2}$H$^{+}$, which have hyperfine components, the fitting was done with multiple Gaussian functions, and the $\Delta$v values used in this analysis correspond to the main components. These values are presented in Table\,\ref{resultsD}. The absence of values in some sources is due to a lack of emission or that the hyperfine components are completely blended with the main component. 
Table\,\ref{deltaprom} presents the average line widths for each kind of source, and with the aim of analyzing their behavior, following the analysis presented in \citet{naiping21}, we present plots that display relations among the measured $\Delta$v in Figure\,\ref{deltav}.   

\begin{figure*}[h!]
    \centering
    \includegraphics[width=7.5cm]{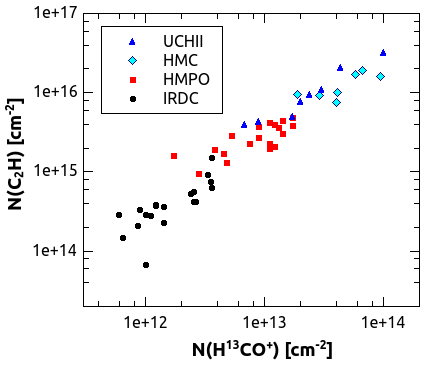}
    \includegraphics[width=7.5cm]{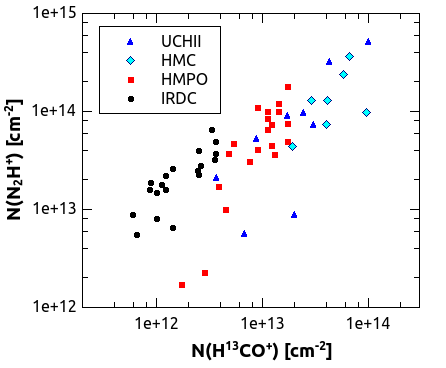}
\caption{Column densities of  C$_{2}$H (left) and N$_{2}$H$^{+}$ (right) against column density of H$^{13}$CO$^{+}$. The column density values were obtained from \citet{gerner14}.}
    \label{denscol}
\end{figure*}

\begin{figure}[h!]
    \centering
    \includegraphics[width=7.5cm]{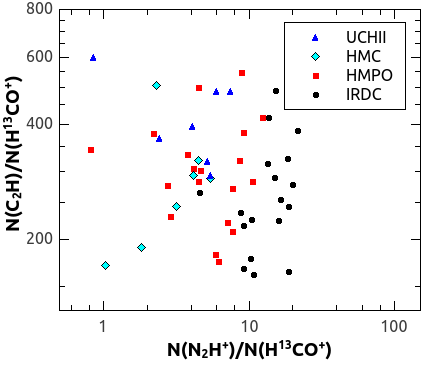}
\caption{Relative abundance [C$_{2}$H]/[H$^{13}$CO$^{+}$] versus [N$_{2}$H$^{+}$]/[H$^{13}$CO$^{+}$]. The column density values
were obtained from \citet{gerner14}.}
    \label{abund}
\end{figure}

From \citet{gerner14} we obtained the column densities of C$_{2}$H, N$_{2}$H$^{+}$, and H$^{13}$CO$^{+}$ for each source. We used the 
column densities obtained as `iteration 1' according to the model used by the authors, which are the values derived 
with the mean temperatures from the best-fit models of `iteration 0'. Values indicated as upper limits are not included in our analysis. 
Figure\,\ref{denscol} displays the column densities of C$_{2}$H and N$_{2}$H$^{+}$ versus the H$^{13}$CO$^{+}$ column density. Error bars are not included, given the authors did not inform them in their work.  Finally, Fig.\,\ref{abund} displays the relative abundance [C$_{2}$H]/[H$^{13}$CO$^{+}$] versus [N$_{2}$H$^{+}$]/[H$^{13}$CO$^{+}$]. 

\section{Discussion}
\label{discus}

Unlike low-mass stars, the formation of high-mass stars is not yet fully understood, and different scenarios 
are proposed (see e.g. \citealt{motte18,tan14}). Probing the chemical conditions of high-mass star-forming regions at different evolutive stages is an important issue to advance in the knowledge on the formation of this kind of stars, and to explore such chemical conditions, it is necessary to use efficient tools. 

In this context, we decided to test the HCN/HNC ratio as a thermometer of the gas, and the use of the emission of H$^{13}$CO$^{+}$, HC$_3$N, N$_{2}$H$^{+}$, and C$_{2}$H as  ``chemical clocks''. The obtained results from each study are discussed in what follows.

\subsection{Testing the use of HCN/HNC ratio as a thermometer}

Employing simple molecules, ubiquitous in the star-forming regions, to derive their physical parameters is a helpful strategy for methodically analyzing many  regions and avoiding complex calculations. In this context, we use the HCN/HNC intensity ratio formulation proposed recently by \citet{hacar20} to estimate the kinetic temperature (T$_{\rm K}$(HCN/HNC)). As done by the authors, we probed them through comparisons with dust temperature (T$_{\rm dust}$) and T$_{\rm K}$ obtained from the ammonia emission ($\rm T_{K}(NH_{3})$), when those parameters were available towards the sources of the analyzed sample. It is worth noting that the useful kinetic temperature range derived from the ammonia is between 15 and 40 K \citep{hot83,urqu11}, in coincidence with the valid range indicated for T$_{\rm K}$(HCN/HNC). 

Despite the limited amount of sources satisfying that the obtained T$_{\rm K}$ from the HCN/HNC ratio is $>$\,15 K with measured values of T$_{\rm dust}$ and T$_{\rm K}$(NH$_{3}$) (see Table\,\ref{tkresults}) some conclusion can be extracted. Also, from the sources with T$_{\rm K}$(HCN/HNC)$<$15 K or lower than expected, we can obtain important conclusions about this tool. 

The comparison between T$_{\rm dust}$ and T$_{\rm K}$(NH$_{3}$) (see upper panel of Figure\,\ref{compara}) indicates  that, within a range of 0--5 K of difference between both temperatures, the molecular gas and the dust could be considered that are thermally coupled, allowing us to suggest that in general this property could be extrapolated to the whole sample. This supports that 
the comparison between $\rm T_{K}$(HCN/HNC), which is obtained from the gas emission, and the T$_{\rm dust}$ can be made.

Valid kinetic temperature values obtained following the HCN/HNC ratio (i.e. $>$15 K) seem to be more accurate for IRDCs and HMPOs than 
for HMCs and UC\hii~regions. By analyzing the average values (see Table\,\ref{deltaT}), $\rm T_{K}$ obtained from the dust and ammonia emissions increases from IRDCs to UC\hii~regions as it is expected, but $\rm T_{K}$(HCN/HNC) increases only from IRDCs to HMPOs, and
then it decreases in HMCs and UC\hii~regions. This suggests that in such sources, using the HCN-HNC tool to derive T$_{\rm K}$ can be not appropriated, and it may underestimate the actual temperature value. 

\citet{hacar20} calculated the T$_{\rm K}$(HCN/HNC) towards a large sample of dense clumps extracted from the MALT90 survey (\citealt{foster11,jackson13}), and compare them with the T$_{\rm dust}$ measurements derived by \citet{guzman15}. At this point, they stated that the comparison between T$_{\rm K}$ and T$_{\rm dust}$ should be done with caution given the fact that the beam resolution of the MALT90 targets and the measurement sensitivity within this beam does not allow to resolve the temperature of individual clumps. Anyway, they pointed out that a correspondence between both temperatures prevails despite the mentioned observational caveats. However, they do not discuss the optical depth issue in the MALT90 sample as they did for their Orion results. This is an important matter to be considered, mainly when we work with a sample of different kinds of sources. As \citet{hacar20} mention, the increase of the HCN J=1--0 line opacity, while the HNC J=1--0 line likely remains optically thin, would reduce the HCN/HNC ratio, decreasing the values of T$_{\rm K}$(HCN/HNC). This phenomenon could be occurring in the sources with T$_{\rm K}$(HCN/HNC)$<$15 K (see Sect.\,\ref{low15}) and
in some other sources that, although the T$_{\rm K}$ obtained from the HCN-HNC tool is greater than 15 K, their values are well below the temperatures obtained from the dust and/or ammonia (about 10 K of difference). This is the case of HMCs and UC\hii~regions of our sample, in which, in general, it is observed that the HCN hyperfine line F=1-1 appears absorbed, indicating high optical depths (for an example, see Fig.\,\ref{spectra5}).

\subsubsection{Sources with T$_{\rm K}$(HCN/HNC)$<$15 K}
\label{low15}

In our sample, there are several sources that the HCN/HNC ratio yields kinetic temperatures lower than 15 K, which were not included
in the analysis nor the comparisons with T$_{\rm dust}$ and T$_{\rm K}$(NH$_{3}$) performed above. 
These sources have I(HCN)/I(HNC) ratios close to the unity or even lower than one, which according to \citet{hacar20}, in those cases, the uncertainties seem to increase, likely due to the combination of excitation and opacity effects. The percentages of sources with this issue is: 58\%, 30\%, 28\%, and 11\% for IRDCs, HMPOs, HMCs, and
UC\hii~regions, respectively, suggesting that may be a condition that is more pronounced at earlier stages.

\begin{figure}
    \centering
    \includegraphics[width=9cm]{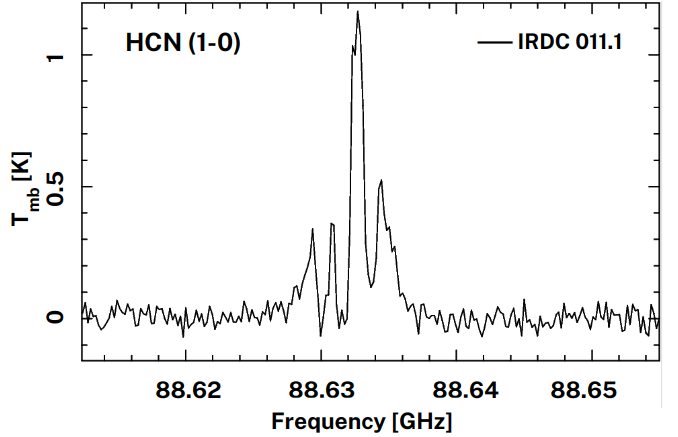}
\caption{Example of an HCN spectrum with strong signatures of saturation which affects the use of HCN-HNC ratio as a thermometer yielding unreliable temperatures below 15 K.}
    \label{hcnspect}
\end{figure}

Taking into consideration the HCN hyperfine line anomalies \citep{wal82,lough12}, which prevent obtaining reliable values of HCN opacities, we carefully analyzed each HCN and HNC spectrum to look for signatures of high optical depths. We found that the HCN spectra of sources with T$_{\rm K}$(HCN/HNC)$<$15 K have pronounced saturation and/or self-absorption features in some cases at both the main component and hyperfine lines (see Fig.\,\ref{hcnspect} for an example), and in other cases in the hyperfine line F=1-1. In the case of the HNC, we did not find such spectral features.

\bigskip

We point out that the use of the HCN/HNC ratio as an universal thermometer in the ISM should be taken with
care. We suggest that such a thermometer could be used only in some IRDCs and HMPOs (when the derived kinetic temperature is not lower than 15 K; see Sect.\,\ref{low15}) and in more evolved regions, for instance, HMCs and UC\hii~regions, this tool underestimates the temperature. 
Thus, we conclude that the HCN-HNC tool as a kinetic temperature estimator should be used 
only after a careful analysis of the HCN spectrum, checking that no line, neither the main nor the hyperfine ones, presents absorption features;
otherwise, it is not an useful tool for calculating kinetic temperatures.

\subsection{An exploration of molecules as ``chemical clocks''}
 
Molecular species such as HNC, HCN, H$^{13}$CO$^{+}$, C$_{2}$H, and HC$_{3}$N are relatively easy to be observed and they 
give us a significant amount of scope to
search for differences in the chemistry as a function of the evolutionary stage in massive star forming regions \citep{sanhueza12,naiping21,urquhart}. 
After a similar analysis as presented by \citet{naiping21} performed to our sample of massive star forming regions, in what follows we discuss our results. 

Firstly, from the comparison between the molecular integrated intensities and the 870 $\mu$m peak flux (see Fig.\,\ref{F870}), we found that, in general, the integrated intensities of N$_{2}$H$^{+}$, C$_{2}$H, HNC, and HC$_{3}$N increase with the submillimeter peak flux. In the case of IRDCs, HMPOs, and HMCs, this increment correlates with the evolutive stage of the sources, i.e. the position of each kind of source in the plots seems to be sectorized. UC\hii~regions show the same trend, but their points are overlapped with those of the other sources in all cases except for the N$_{2}$H$^{+}$ case. \citet{naiping21} found that the N$_{2}$H$^{+}$ integrated intensities of \hii~regions tend to be lower that those for MYSOs. This behavior seems to occur with the UC\hii~regions of our sample in comparison with the others sources, suggesting that the destruction path due to electronic recombination, where N$_{2}$H$^{+}$ is destroyed by free electrons from the surroundings \citep{vigren}, could be ongoing in the UC\hii~regions. This result complements what was presented by \citet{naiping21} with sources at earlier stages, and suggests that the relation between the molecular integrated intensities and the 870 $\mu$m peak flux could be useful to distinguish regions among IRDCs, HMPOs, and HMCs.

The comparisons between the column densities obtained from \citet{gerner14} show also an interesting trend regarding the evolutive stage of the sources. \citet{naiping21} stated, based on astrochemical models (e.g. \citealt{bergin97,nomura04}), that the H$^{13}$CO$^{+}$ column density reflects the amount of H$_{2}$ density in a clump because its abundance seems to not vary much with the time. In line with their work, we compared N(C$_{2}$H) and N(N$_{2}$H$^{+}$) with N(H$^{13}$CO$^{+}$), 
(see Fig.\,\ref{denscol}). 
The N(H$^{13}$CO$^{+}$) values of our sample range in a wider interval of values (from some 10$^{12}$ cm$^{-2}$ for IRDCs to values close to 10$^{14}$ cm$^{-2}$ for HMCs) than the N(H$^{13}$CO$^{+}$) presented by \citet{naiping21}. An increase in the N(C$_{2}$H) and N(N$_{2}$H$^{+}$) with the N(H$^{13}$CO$^{+}$) is observed. In the case of N(C$_{2}$H) versus N(H$^{13}$CO$^{+}$) relation, such an increment also presents a conspicuous correlation with the evolutive stage from IRDCs to HMCs in which the positions of each kind of source is sectorized in the plot. In the UC\hii~regions, both column densities have similar values compared with those of HMCs and the largest ones of HMPOs. A similar behavior, with larger dispersion, is observed in the comparison between N(N$_{2}$H$^{+}$) versus N(H$^{13}$CO$^{+}$). 
The N$_{2}$H$^{+}$ and H$^{13}$CO$^{+}$ column densities increment from IRDCs to HMCs could be explained by their progressive formation through the H$_{3}^{+}$ mechanisms ($\rm H_{3}^{+} +$ $^{13}$CO $\rightarrow \rm H^{13}CO^{+} + H_{2}$, and $\rm H_{3}^{+} + N_{2} \rightarrow \rm N_{2}H^{+} + H_{2}$; \citealt{jorgensen04}).
Then, the brake on the constant increment in the column density values in the UC\hii~regions compared with the previous sources can be due to the beginning of the destruction of such molecules by electronic recombination \citep{naiping16}.

It is worth noting that the discussed behavior in the column densities (which have molecular excitation assumptions) regarding the correlation with the evolutive stage from IRDCs to HMCs is quite similar to the comparison between the integrated intensities and the peak submillimeter flux (which are direct measurements).  
Thus, in the line of a chemical clock analysis, even though it is necessary more statistics, we propose that these relations, mainly the N(N$_{2}$H$^{+}$) versus N(H$^{13}$CO$^{+}$) and the I(C$_{2}$H) versus F$_{870 {\rm \mu m}}$, in which the different kinds of sources (IRDCs, HMPOs, and HMCs) are clearly separated, could be used to differentiate them, which can be useful in works handling with a large amount of sources of unknown, or not completed known, nature.

The comparison between the relative abundance ratios [C$_{2}$H]/[H$^{13}$CO$^{+}$] versus [N$_{2}$H$^{+}$]/[H$^{13}$CO$^{+}$] 
(Fig.\,\ref{abund}) presents a high dispersion, mainly along the y-axis, i.e. in the N(C$_{2}$H)/N(H$^{13}$CO$^{+}$) ratio, and does not show a remarkable difference as it was found between MYSOs and \hii~regions by \citet{naiping21}.  However, along the x-axis, i.e. in the 
N(N$_{2}$H$^{+}$)/N(H$^{13}$CO$^{+}$) ratio, some trend can be appreciated, mainly such a ratio 
seems to be larger in IRDCs in comparison with the other kind of sources, which in the line of a chemical clock analysis, we suggest that this abundance ratio can be used to differentiate the earliest stage of the star forming regions.

The molecular line widths (FWHM) are usually associated with the kinematics of the molecular gas and can be affected by multiple events, such as shocks, outflows, rotation of the clump, and turbulence. In this work, we assume that line widths originate mainly from the increasing turbulence that arises as a consequence of evolving star formation stages \citep{sanhueza12,naiping21}. With this in mind, by analyzing the average line width of N$_{2}$H$^{+}$, HC$_{3}$N, H$^{13}$CO$^{+}$ and C$_{2}$H (see Table\,\ref{deltaprom}), it can be appreciated that in the four species, the line width rises until reaching the HMC stage. 
Such an increase could be a consequence of the gas dynamics related to the star-forming processes that take place in the molecular clumps \citep{fontani02,pigo03}.  
Particularly, our $\Delta$v(N$_{2}$H$^{+}$) average values are in quite agreement with those presented 
by \citet{sanhueza12}: the $\Delta$v(N$_{2}$H$^{+}$) average value that we obtained for IRDCs is similar to that obtained by the authors in their so-called ``quiescent'' and ``intermediate'' clumps, while the value that we obtained for HMPOs is in agreement with the values of their ``active'' clumps. Finally, our $\Delta$v(N$_{2}$H$^{+}$) average values obtained for HMCs and UC\hii~regions are 
slightly larger than the value obtained for their ``red'' clumps. Following \citet{sanhueza12}, we can confirm that line widths of N$_{2}$H$^{+}$ slightly increase with the evolution of the clumps. 

As found by \citet{sanhueza12} in their sample of clumps embedded in IRDCs, we observed that in our sources, the H$^{13}$CO$^{+}$ and HC$_{3}$N have slightly narrower line widths than N$_{2}$H$^{+}$ (see panels d and f in Fig.\,\ref{deltav}). This hints that they trace similar optically thin gas emanating from the internal layers of the regions, and according to our findings, it seems to be independent of the kind of source. 
\citet{naiping21} found that the best $\Delta$v correlation is between C$_{2}$H and N$_{2}$H$^{+}$ for their sample of EGOs. In our case, we did not find such a well correlation between these molecular species (see panel e in Fig.\,\ref{deltav}). The C$_{2}$H line widths are slightly narrower than those of the N$_{2}$H$^{+}$. \citet{sanhueza12} also remarked that the C$_{2}$H appears to present the best correlation with N$_{2}$H$^{+}$ among all the studied $\Delta$v relations. However, by inspecting their $\Delta$v(C$_{2}$H) versus $\Delta$v(N$_{2}$H$^{+}$) plot, we observe a similar behavior as our plot, where most of the points tend to be below the unity line. 

In the spectra of these four molecular species towards the whole sample, we did not find line wings that may probe outflows, except for the HC$_{3}$N spectra of HMPO18247, HMC029.96 (see Fig.\,\ref{spectra6}), UCH10.30, and in the H$^{13}$CO$^{+}$ spectrum of HMC34.26, in which small spectral wings appear. UCH10.30 is an EGO; the other sources present 4.5 $\mu$m extended emission and diffuse emission at the Ks band. Besides these sources, it can be noticed that the near-IR evidence of jets or/and outflows have not any direct correlation with spectral features in the emission of N$_{2}$H$^{+}$, HC$_{3}$N, H$^{13}$CO$^{+}$, and C$_{2}$H.

\section{Summary and concluding remark}
\label{concl}

We presented a spectroscopic molecular line analysis of 55 high-mass star-forming regions with the aim of probing some chemical tools that were recently proposed to characterize such regions. The analyzed sources are classified as IRDCs, HMPOs, HMCs, and UC\hii~regions, according to an evolutionary progression in the high-mass star formation. The main results can be summarized as follows:

1. The emissions of the HCN and HNC isomers were used to estimate the kinetic temperature through a new `thermometer' whose formulation was proposed by \citet{hacar20}, and we investigated its use in the presented sample of sources. By comparing the T$_{\rm K}$ derived from the HCN/HNC ratio with temperatures obtained from the dust and ammonia emission, we found that the use of such a ratio as a universal thermometer in the ISM should be taken with
care. The HCN optical depth is a big issue to be taken into account. Line saturation may explain the derived temperature values below 15 K, which must be discarded, and mainly correspond to IRDCs and HMPOs. In addition, we found that, although the T$_{\rm K}$ obtained from the HCN-HNC tool is greater than 15 K, their values could be far from the temperatures obtained from the dust or ammonia, yielding lower temperatures. This is the case of HMCs and UC\hii~regions of our sample, in which, in general, it is observed that the HCN hyperfine line F=1-1 appears absorbed.
In conclusion, we point out that the HCN-HNC tool as a kinetic temperature estimator should be used 
only after a careful analysis of the HCN spectrum, checking  that no line, neither the main nor the hyperfine ones, presents absorption features.

2.  Additionally, we have analyzed the molecular species N$_{2}$H$^{+}$, HC$_{3}$N, H$^{13}$CO$^{+}$ and C$_{2}$H in the sample of sources focusing in the use of them as ``chemical clocks''. The comparison of the molecular integrated intensities of HC$_{3}$N, N$_{2}$H$^{+}$, C$_{2}$H, and also of HNC with the 870 $\mu$m peak flux, as well the relations of the column densities of C$_{2}$H and N$_{2}$H$^{+}$ with that of the H$^{13}$CO$^{+}$, can be useful to distinguish regions among IRDCs, HMPOs, and HMCs, which complement the \citet{naiping21} results. On the other hand, from the [C$_{2}$H]/[H$^{13}$CO$^{+}$] versus [N$_{2}$H$^{+}$]/[H$^{13}$CO$^{+}$] relation, we can point out that even though the [C$_{2}$H]/[H$^{13}$CO$^{+}$] ratio shows a large dispersion along the analyzed sources, the [N$_{2}$H$^{+}$]/[H$^{13}$CO$^{+}$] ratio seems to be larger in IRDCs in comparison with the other kind of sources analyzed in this work. 
Finally, regarding the molecular line widths, we found that the $\Delta$v rises from the IRDC to the HMC stage, which could be due to the increasing turbulence as a consequence of the evolution of the star-forming processes. 

In conclusion, this work explores chemical tools, some of them based on direct measurements, to be applied in different ISM environments. In that sense, we probed such tools in a new sample of sources
with respect to previous works, and these results not only contribute to more statistics in the literature but also complement such works with other types of sources.  Using direct tools like, for instance, the ratios of different molecular parameters that can be directly measured from the observations, can be very useful mainly when a large sample of sources is handled. If it is proven that such tools are reliable, they can be used to obtain important statistical information in a simple way. Our work points to it and encourages to perform similar works in larger samples of sources of different types.

\section*{Acknowledgements}

We thank the anonymous referee for her/his useful comments pointing to improve our work. N.C.M. is a doctoral fellow of CONICET, Argentina. S.P. is a member of the {\sl Carrera del Investigador Cient\'\i fico} of CONICET, Argentina. 
This work was partially supported by the Argentina grant PIP 2021 11220200100012 from CONICET.

\bibliography{ms2023-0280ref}{}
\bibliographystyle{aasjournal}



\eject

\appendix
\section{Example spectra}
\label{appx}

\begin{figure}[h!]
    \centering
    \includegraphics[width=15cm]{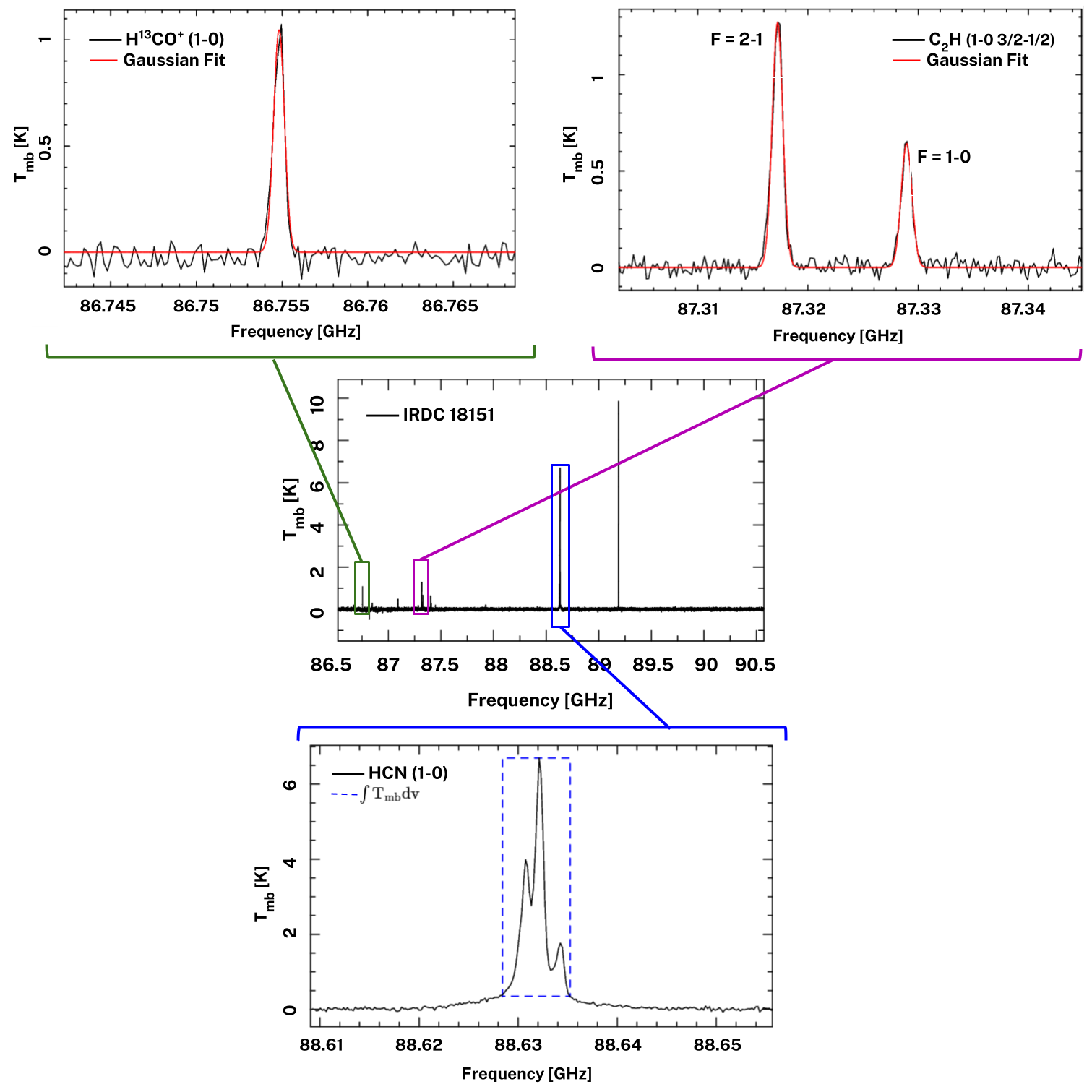}
    \caption{IRAM spectra containg the HCN, H$^{13}$CO$^{+}$, and C$_{2}$H lines towards IRDC 18151. It is shown the integration area in the HCN emission and the Gaussian fittings for the other molecular lines.   }
    \label{spectra1}
\end{figure}

\begin{figure}
    \centering
    \includegraphics[width=16cm]{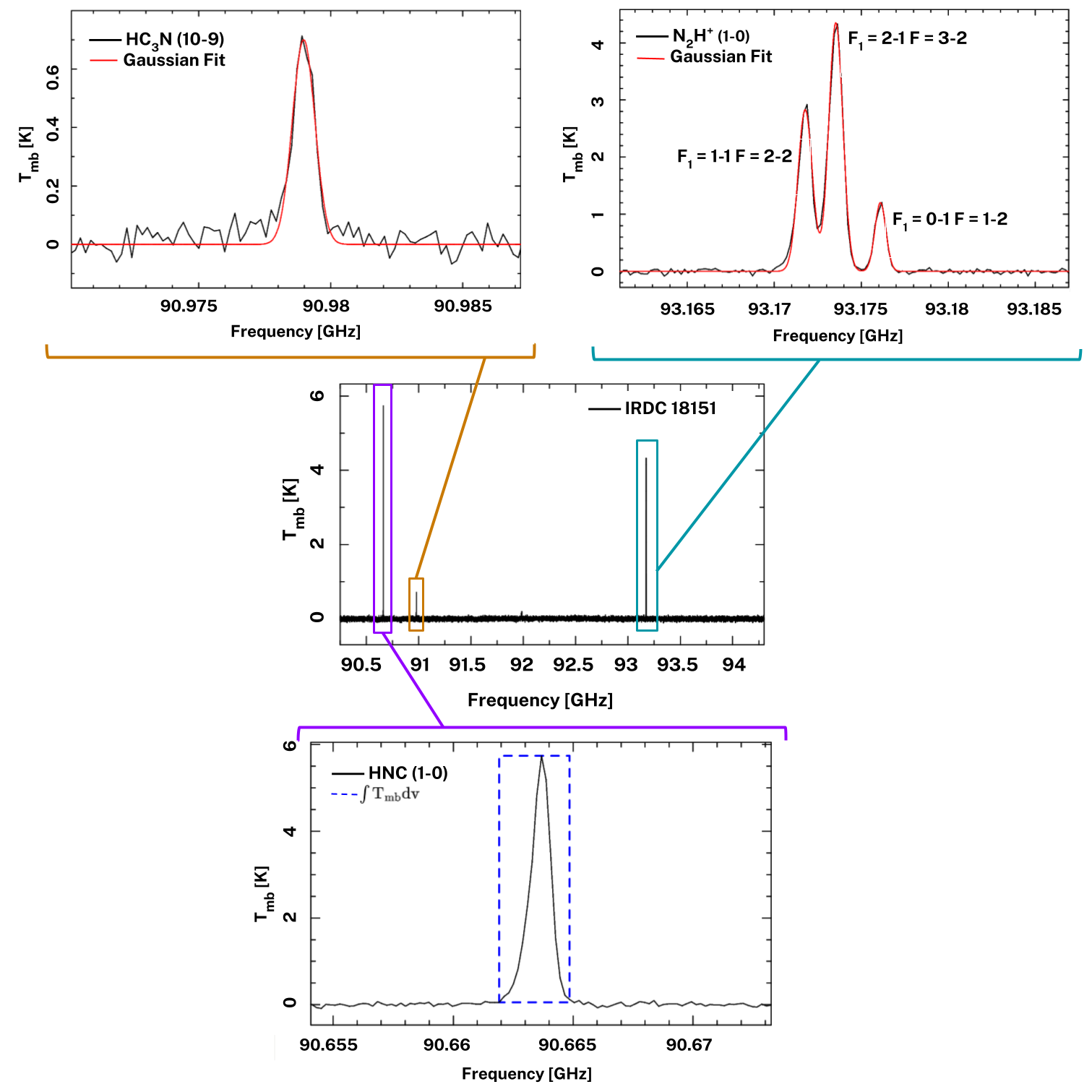}
\caption{IRAM spectra containg the HNC, HC$_{3}$N, and N$_{2}$H$^{+}$ lines towards IRDC 18151. It is shown the integration area in the HNC emission and the Gaussian fittings for the other molecular lines.  }
    \label{spectra2}
\end{figure}

\begin{figure}
    \centering
    \includegraphics[width=16cm]{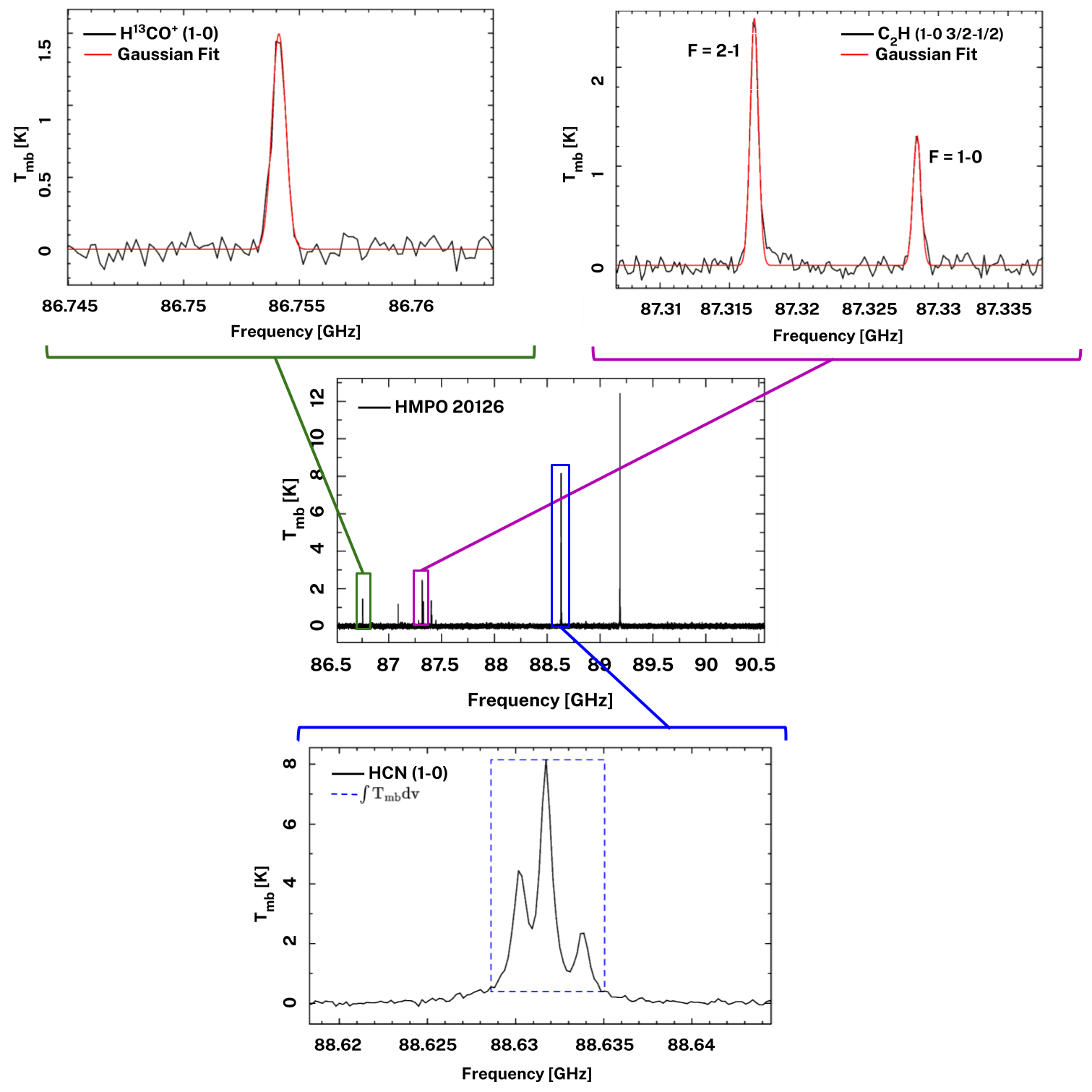}
    \caption{IRAM spectra containg the HCN, H$^{13}$CO$^{+}$, and C$_{2}$H lines towards HMPO 20126. It is shown the integration area in the HCN emission and the Gaussian fittings for the other molecular lines.   }
    \label{spectra3}
\end{figure}

\begin{figure}
    \centering
    \includegraphics[width=16cm]{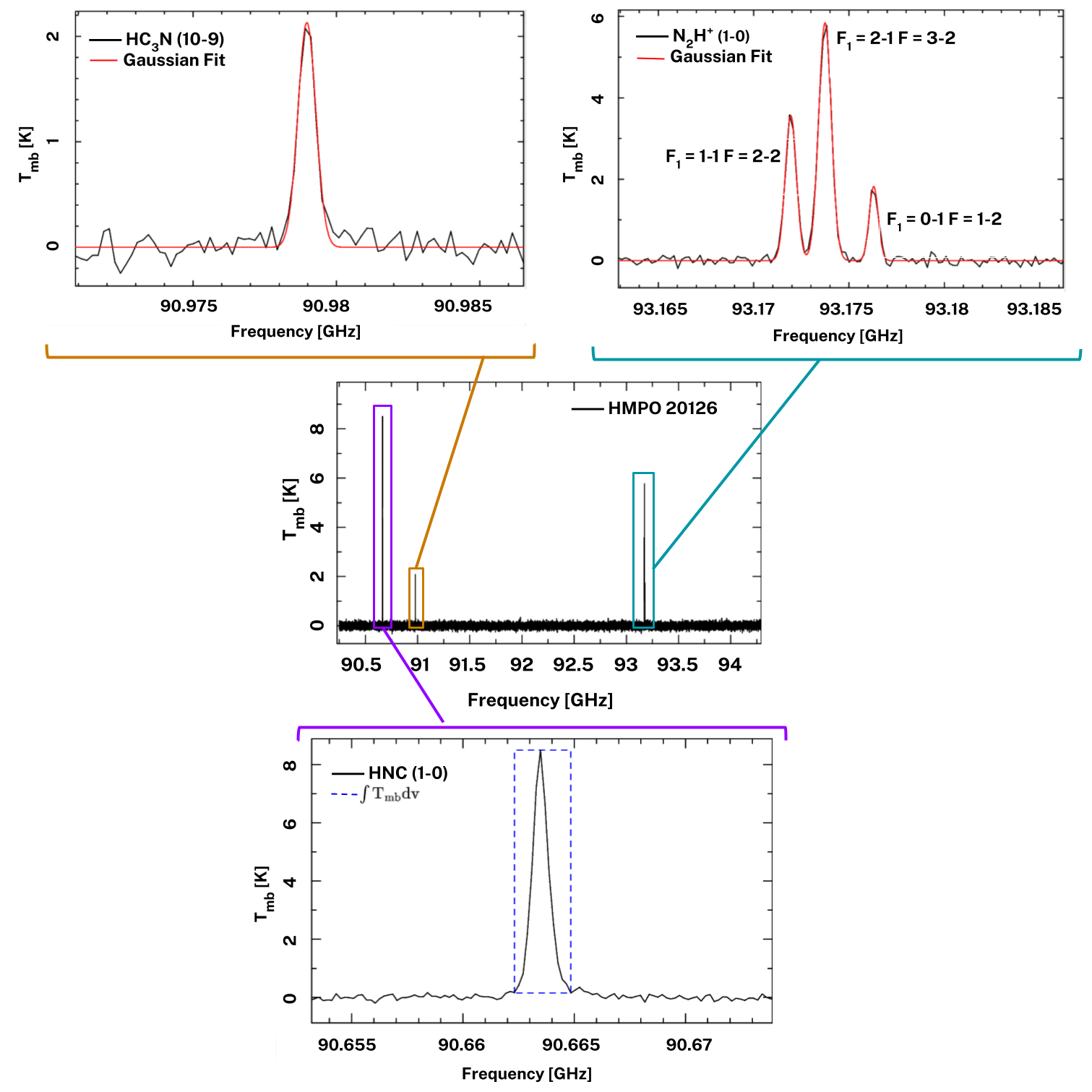}
\caption{IRAM spectra containg the HNC, HC$_{3}$N, and N$_{2}$H$^{+}$ lines towards HMPO 20126. It is shown the integration area in the HNC emission and the Gaussian fittings for the other molecular lines.  }
    \label{spectra4}
\end{figure}

\begin{figure}
    \centering
    \includegraphics[width=16cm]{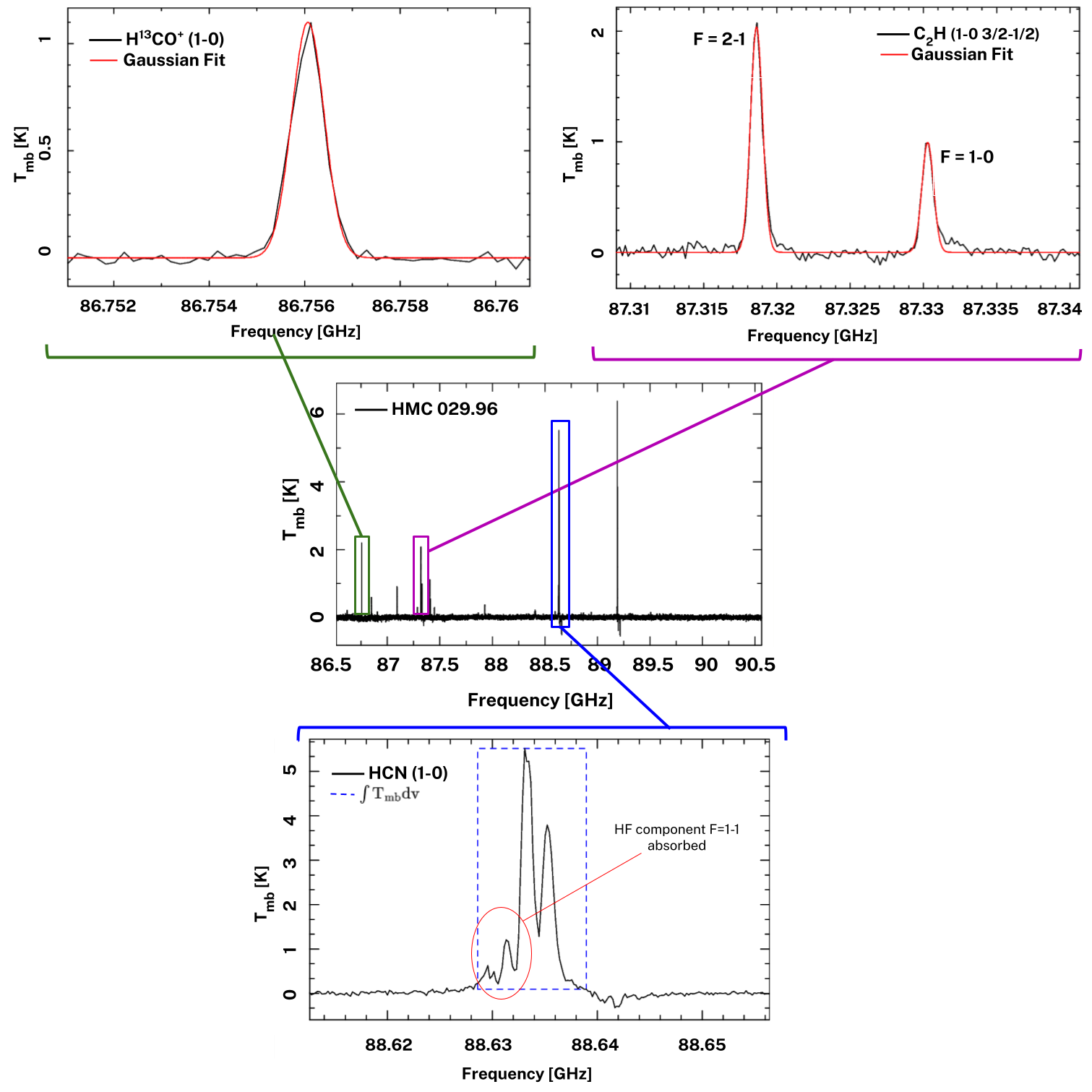}
    \caption{IRAM spectra containg the HCN, H$^{13}$CO$^{+}$, and C$_{2}$H lines towards HMC 029.96. It is shown the integration area in the HCN emission and the Gaussian fittings for the other molecular lines. In the case of the HCN it is remarked that the hyperfine component F=1-1 is absorbed.}
    \label{spectra5}
\end{figure}

\begin{figure}
    \centering
    \includegraphics[width=16cm]{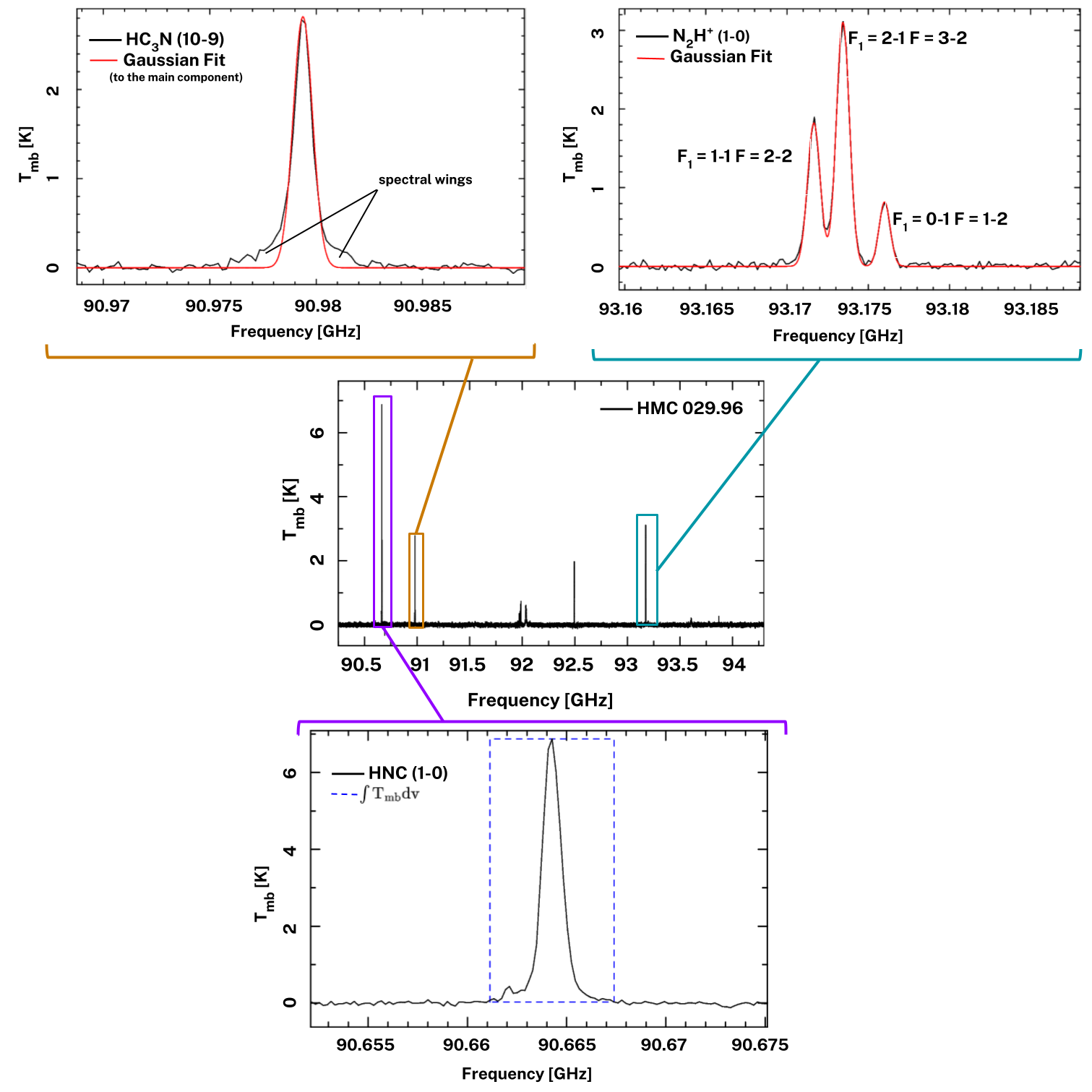}
\caption{IRAM spectra containg the HNC, HC$_{3}$N, and N$_{2}$H$^{+}$ lines towards HMC 029.96. It is shown the integration area in the HNC emission and the Gaussian fittings for the other molecular lines. In the case of the HC$_{3}$N emission small spectral wings appear, and the gaussian fitting corresponds to the main component. }
    \label{spectra6}
\end{figure}

\begin{figure}
    \centering
    \includegraphics[width=16cm]{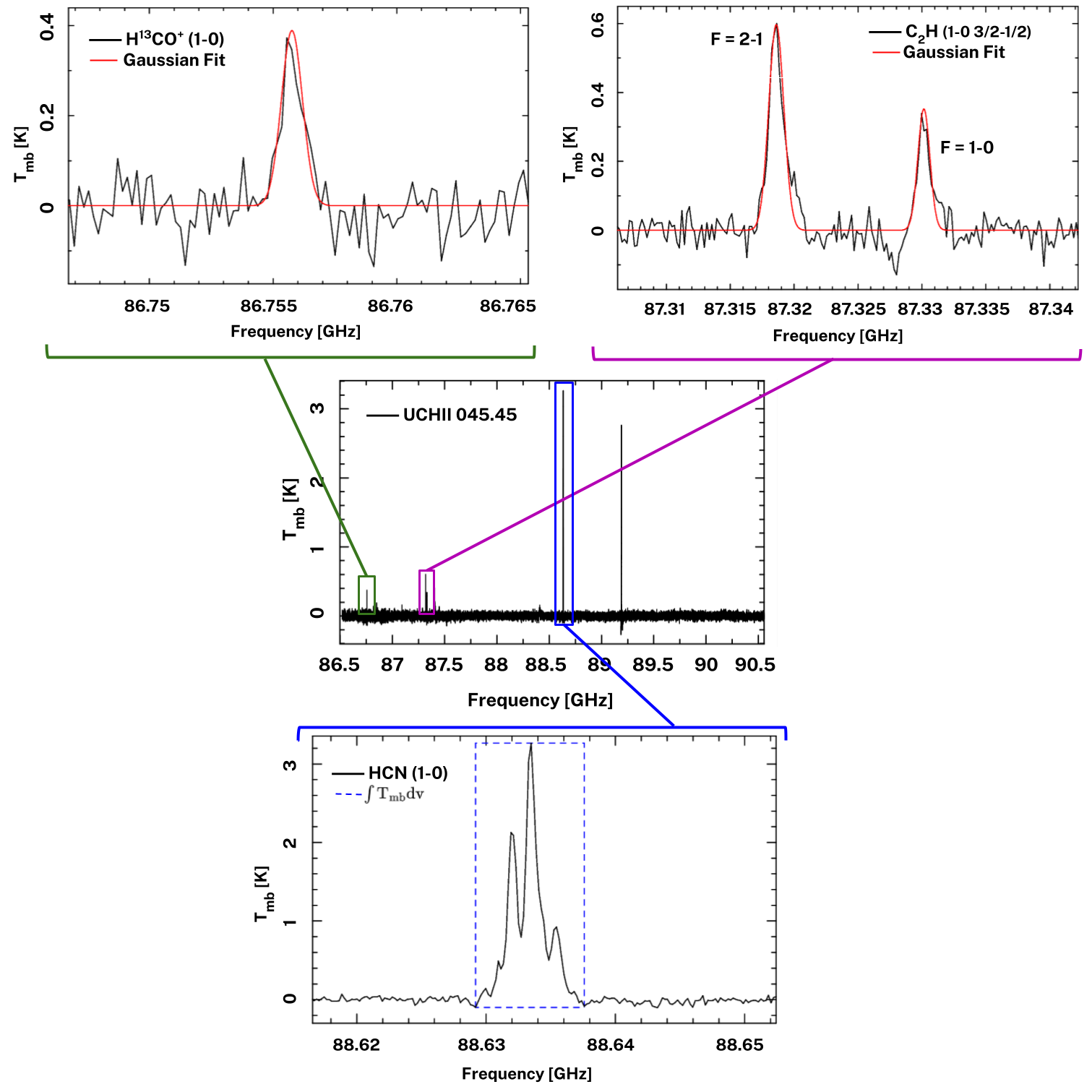}
    \caption{IRAM spectra containg the HCN, H$^{13}$CO$^{+}$, and C$_{2}$H lines towards UCHII 45.45. It is shown the integration area in the HCN emission and the Gaussian fittings for the other molecular lines.   }
    \label{spectra7}
\end{figure}

\begin{figure}
    \centering
    \includegraphics[width=16cm]{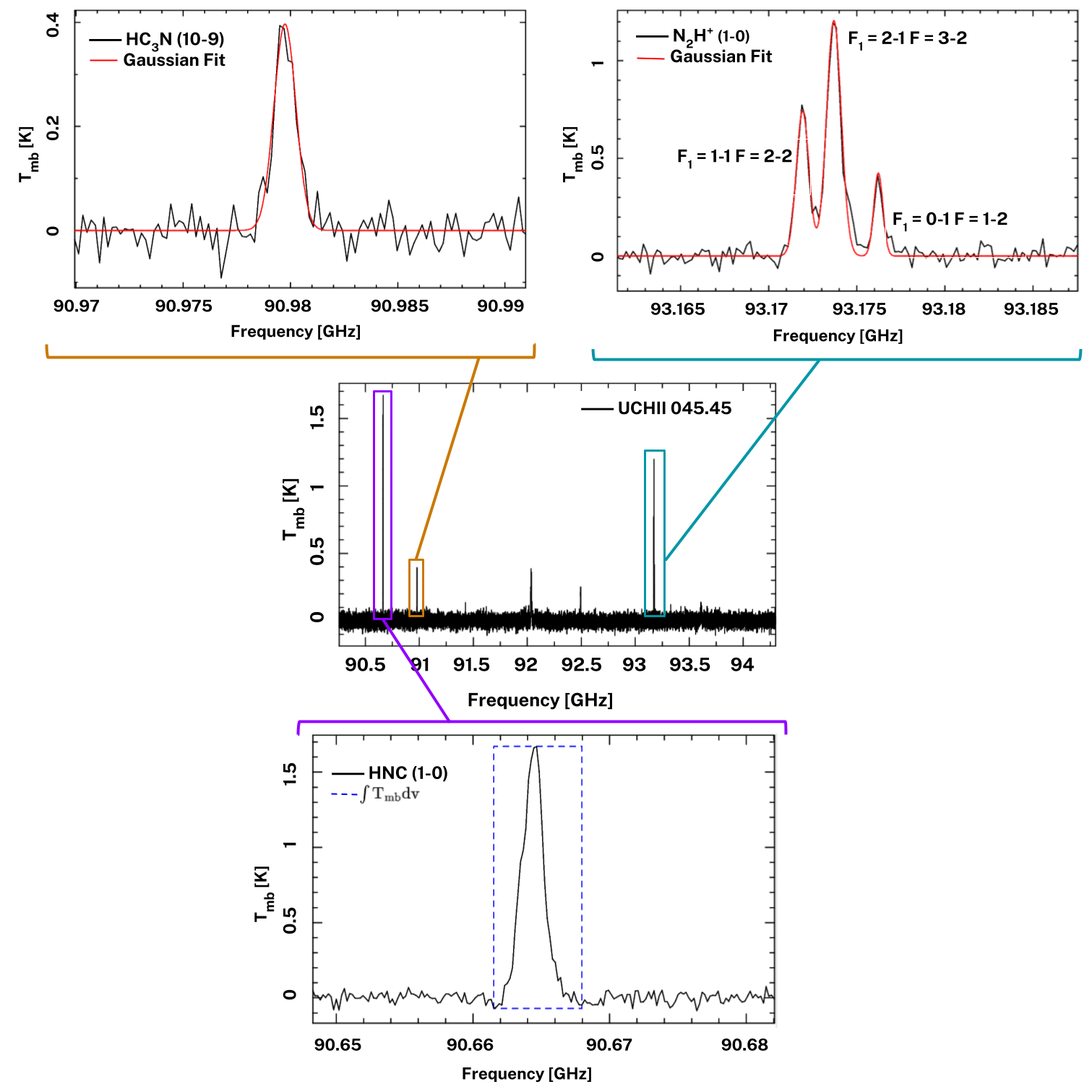}
\caption{IRAM spectra containg the HNC, HC$_{3}$N, and N$_{2}$H$^{+}$ lines towards UCHII 45.45. It is shown the integration area in the HNC emission and the Gaussian fittings for the other molecular lines.  }
    \label{spectra8}
\end{figure}

In this appendix we include some spectra of the analyzed molecular lines as an example of each kind of source, showing the integration area in the case of 
the HCN and HNC emissions and the Gaussian fittings for the H$^{13}$CO$^{+}$, C$_{2}$H, HC$_{3}$N, and N$_{2}$H$^{+}$. Spectra of IRDC 18151 are shown in Fig.\ref{spectra1} and \ref{spectra2},  of HMPO 20126 in Figs.\,\ref{spectra3} and \ref{spectra4}, of HMC 0.29.26 in Figs.\,\ref{spectra5} and \ref{spectra6}, and of UCHII 45.45 in Figs.\,\ref{spectra7} and \ref{spectra8}.

\end{document}